\documentclass[%twocolumn,
preprintnumbers,superscriptaddress
,footinbib
]{revtex4-1}

\usepackage{simplewick}

\newcommand{\beq}{\begin{eqnarray}}
\newcommand{\eeq}{\end{eqnarray}}

\newcommand{\nn}{\nonumber\\}

\newcommand{\rcrit}{r_{\rm crit}}

\usepackage{graphicx}
\usepackage{amssymb}
\usepackage{amsmath}
\usepackage{color}
\usepackage{epsfig}
\usepackage{epstopdf}
\usepackage{hyperref}
\usepackage{subfigure}
\usepackage{comment}
\usepackage{bm}
\usepackage{ulem}
\usepackage{url}
\allowdisplaybreaks

\usepackage{mathtools}

\DeclarePairedDelimiterX\braket[2]{\langle}{\rangle}{#1 \delimsize\vert #2}

\begin{document}

\preprint{RIKEN-iTHEMS-Report-24, YITP-24-73}

%\date{\today}
\title{
Scale setting and hadronic properties in light quark sector with $(2+1)$-flavor Wilson fermions at the physical point
}

\author{Tatsumi~Aoyama}
%\email[]{}
\affiliation{Institute for Solid State Physics, University of Tokyo, Chiba 277-8581, Japan}

\author{Takahiro~M.~Doi}
%\email[]{doi.takahiro.5d@kyoto-u.ac.jp}
\affiliation{Department of Physics, Kyoto University, Kitashirakawa Oiwakecho, Sakyo-ku, Kyoto, 606-8502, Japan}
\affiliation{Interdisciplinary Theoretical and Mathematical Sciences Program (iTHEMS), RIKEN, Wako 351-0198, Japan}

\author{Takumi~Doi}
%\email{doi@ribf.riken.jp}
\affiliation{Interdisciplinary Theoretical and Mathematical Sciences Program (iTHEMS), RIKEN, Wako 351-0198, Japan}

\author{Etsuko~Itou}
\email[]{itou@yukawa.kyoto-u.ac.jp}
\affiliation{Center for Gravitational Physics, Yukawa Institute for Theoretical Physics, Kyoto University, Kitashirakawa Oiwakecho, Sakyo-ku, Kyoto 606-8502, Japan}
\affiliation{Interdisciplinary Theoretical and Mathematical Sciences Program (iTHEMS), RIKEN, Wako 351-0198, Japan}
%\affiliation{\it {Department of Physics, and Research and Education Center for Natural Sciences, Keio University, 4-1-1 Hiyoshi, Yokohama, Kanagawa 223-8521, Japan}}
%\affiliation{\it {Research Center for Nuclear Physics (RCNP), Osaka University, Osaka 567-0047, Japan}}

\author{Yan~Lyu}
%\email[]{}
\affiliation{Interdisciplinary Theoretical and Mathematical Sciences Program (iTHEMS), RIKEN, Wako 351-0198, Japan}

\author{Kotaro~Murakami}
%\email{kotaro.murakami@yukawa.kyoto-u.ac.jp}
\affiliation{Department of Physics, Tokyo Institute of Technology, 2-12-1 Ookayama, Megro, Tokyo 152-8551, Japan}
\affiliation{Interdisciplinary Theoretical and Mathematical Sciences Program (iTHEMS), RIKEN, Wako 351-0198, Japan}

\author{Takuya~Sugiura}
%\email[]{sugiura@rcnp.osaka-u.ac.jp}
\affiliation{Faculty of Data Science, Rissho University, Kumagaya 360-0194, Japan}
\affiliation{Interdisciplinary Theoretical and Mathematical Sciences Program (iTHEMS), RIKEN, Wako 351-0198, Japan}

%\author{...}
%\email[]{}
%\affiliation{
%}

\collaboration{HAL QCD Collaboration}

%%%%%%%%%%%%%%%%%%%%%%%%%%%%%%%%%%%%%%
\begin{abstract}

We report scale setting and
hadronic properties 
for our new lattice QCD gauge configuration set (HAL-conf-2023).
We employ $(2+1)$-flavor nonperturbatively improved Wilson fermions with stout smearing
and the Iwasaki gauge action on a $96^4$ lattice,
and generate configurations of 8,000 trajectories at the physical point.
We show the basic properties of the configurations
such as the plaquette value, topological charge distribution and their auto-correlation times.
The scale setting is performed by detailed analyses of the $\Omega$ baryon mass.
We calculate the physical results of quark masses,
decay constants of pseudoscalar mesons and single hadron spectra in light quark sector.
The masses of the stable hadrons are found to agree with the experimental values 
within a sub-percent level.

\end{abstract}

\maketitle

\renewcommand{\thefootnote}{\arabic{footnote}}
\setcounter{footnote}{0}

\newpage

%\tableofcontents %To be removed for PRL submission

\setcounter{page}{1}

%%%%%%%%%%%%%%%%%%%%%%%%%%%%%%%%%%%%%%
%%%%%%%%%%%%%%%%%%%%%%%%%%%%%%%%%%%%%%
\section{Introduction}
%%%%%%%%%%%%%%%%%%%%%%%%%%%%%%%%%%%%%%
%%%%%%%%%%%%%%%%%%%%%%%%%%%%%%%%%%%%%%

Quantum Chromodynamics (QCD) governs the dynamics of quarks and gluons,
and ultimately properties of hadrons and nuclei as well as
nuclear astrophysical phenomena such as binary neutron star merges and nucleosynthesis.
Currently, numerical lattice QCD simulation is the only available theoretical method
which can solve QCD in a first-principle manner,
and decades of theoretical progress
together with the development of faster supercomputers
enable us to make predictions (or postdictions)
for basic hadronic quantities
as well as quantities relevant to the search of
physics beyond the Standard Model~\cite{FlavourLatticeAveragingGroupFLAG:2021npn, Aoyama:2020ynm, Borsanyi:2020mff}.

It remains, however, a great challenge to
construct a bridge between QCD and nuclear physics
by lattice QCD, since it is necessary to study
multi-hadron systems on the lattice.
For such studies, the so-called Signal(S)-to-Noise(N) problem appears generally,
where the S/N ratio of the relevant correlation function 
becomes exponentially worse
with lighter quark masses, a larger mass number and
larger Euclidean time separation between source and sink operators~\cite{Parisi:1983ae, Lepage:1989hd}.
In addition,
it is necessary to employ a larger lattice volume
compared to that required for usual single hadron studies,
in order to accommodate the longest interaction range
(typically the range of the one-pion exchange) between two hadrons.
It is clear that these problems are most pronounced
with the physical (light) quark masses,
while it is expected that 
physical point simulation is particularly important for nuclear physics
due to the vital role of one- (and two-) pion exchanges.

The HAL QCD Collaboration significantly improved the situation recently,
where we performed the first lattice QCD calculations for
baryon-baryon interactions near the physical point~\cite{Gongyo:2017fjb, HALQCD:2018qyu, HALQCD:2019wsz, Lyu:2021qsh, Aoki:2020bew}.
(See also Refs.~\cite{Francis:2018qch, Horz:2020zvv, Amarasinghe:2021lqa, Green:2021qol} for recent studies for two-baryon systems at heavy quark masses
based on L\"uscher's finite volume method.)
Utilizing the (time-dependent) HAL QCD method~\cite{Ishii:2006ec, Ishii:2012ssm},
which can significantly ameliorate the S/N problem~\cite{Ishii:2012ssm, Iritani:2018vfn}, 
we performed comprehensive calculations of baryon-baryon interactions relevant to nuclear physics~\cite{Aoki:2020bew}
as well as meson-meson~\cite{Lyu:2023xro} and meson-baryon interactions~\cite{Lyu:2022imf} 
at $(m_\pi, m_K) = (146, 525)$ MeV on a large volume of $L^4 = (8.1 {\rm fm})^4$
with $N_s \equiv L/a=96$ and a lattice cutoff 
%$a^{-1} = 2333(18)$ MeV [XXX].
$a^{-1} = 2.3$ GeV~\cite{Ishikawa:2015rho}.
In these calculations, 
we employed the gauge configurations 
with $\sim$ 2,000 trajectories
generated by the PACS Collaboration
on the K computer, one of the fastest supercomputers in the world of that time.
(We refer to this configuration set as ``K-conf'' hereafter.)
Some of the highlights are the
prediction of bound/virtual di-baryon states as
$\Omega\Omega$~\cite{Gongyo:2017fjb}, 
$\Omega_{ccc}\Omega_{ccc}$~\cite{Lyu:2021qsh},
$N\Omega$~\cite{HALQCD:2018qyu},
$N\Xi$ (remnant of H-dibaryon)~\cite{HALQCD:2019wsz, Kamiya:2021hdb}
as well as $\Xi$-hypernuclei~\cite{Hiyama:2019kpw, Hiyama:2022jqh} and 
the exotic $T^+_{cc}$ state~\cite{Lyu:2023xro}.

Under these circumstances,
the most important next step is to perform physical point simulations for nuclear physics.
The results of di-baryons/hypernuclei/exotics mentioned above
also indicate its importance.
In fact, these states are found to be located near the unitary regime,
and the importance of one- (or two-) pion exchange is explicitly observed in many of these states.
Therefore, a tiny difference in quark masses could make a significant impact on
the fate of these states.
It is also implied that precision calculations with larger statistics are necessary.

In this paper,
we report the generation of gauge configurations at the physical point
by the HAL QCD Collaboration.
We employ $(2+1)$-flavor nonperturbatively ${\cal O}(a)$-improved Wilson quark action with the stout smearing
and the Iwasaki gauge action at $\beta = 1.82$ on a $N_s^4=96^4$ 
box,
as was the case for the K-conf.
We, however, perform the simulation with the physical-point quark masses, which correspond to
$(m_\pi, m_K) = (
137.1(0.3)(^{+0.0}_{-0.2}),
501.8(0.3)(^{+0.0}_{-0.7})
)$ MeV.
The lattice cutoff, 
$a^{-1} = 2338.8(1.5)(^{+0.2}_{-3.0})$
MeV, 
is also determined in this paper using the mass of the $\Omega$ baryon.
Physical values of
quark masses, decay constants of pseudoscalar mesons
and spectra of other hadrons are evaluated accordingly.

Numerical computations are performed on the supercomputer Fugaku,
the new flagship supercomputer in Japan which succeeds the K computer.
Fugaku was developed by the co-design of hardware and software,
in which lattice QCD was one of the major targets of software applications~\cite{Ishikawa:2021iqw}.
This enables efficient computations of the gauge configuration generation on Fugaku.

Before going to the main part, we have two remarks on our simulations.
First, similar physical point generation was performed by the PACS Collaboration recently,
with the same action and the same 
$(\beta, \kappa_{ud},\kappa_s)$
but with two different volumes, $N_s^4= 64^4$ and $128^4$,
the latter of which is called ``PACS10'' configurations~\cite{Ishikawa:2018jee, PACS:2019ofv}.
With their  determination of the cutoff, $a^{-1} = 2316.2(4.4)$ MeV,
the generation point for PACS10 configurations is found to be $(m_\pi, m_K) = (135.3(6)(3), 497.2(2)(9))$ MeV.
Unfortunately, the statistics are as small as $200$ trajectories
and are insufficient to perform lattice QCD studies for nuclear physics.
In this work, we generate $8,000$ trajectories in total with the benefit from their parameter tuning for the physical point.
We also note that there is worldwide effort to generate configurations
with various lattice QCD actions
near or at the physical point%, see Refs.
~\cite{PACS-CS:2008bkb, PACS-CS:2009sof, BMW:2008jgk, BMW:2014pzb, Ishikawa:2018jee, PACS:2019ofv, RBC:2012cbl, RBC:2014ntl, Bruno:2014jqa, RQCD:2022xux, Alexandrou:2018egz, MILC:2012znn, Bazavov:2017lyh}.

Another remark is on the notion of ``physical point'' in our simulations.
Since we perform $(2+1)$-flavor lattice QCD simulations without QED,
the physical point can be defined only up to the uncertainties of isospin breaking effect
(as well as much minor effect from dynamical heavy quarks).
Since our primary target is to study nuclear physics from lattice QCD,
in particular from the point of view of hadron interactions,
we set our target values for the physical point by employing the isospin-averaged values of pseudoscalar meson masses,
$(m_\pi, m_K) = (m_{\pi^{\rm ave}}, m_{K^{\rm ave}}) = (138, 496)$ MeV.

This paper is organized as follows. 
In Sec.~\ref{sec:setup}, we give simulation details. In Sec.~\ref{sec:property}, we show the basic properties of the generated ensemble by studying the plaquette value, 
the topological charge distribution, 
and their autocorrelation times.
The analyses of the correlation functions of mesons and baryons
are presented in Secs.~\ref{sec:meson-corr} and ~\ref{sec:baryon-corr}, respectively.
In Sec.~\ref{sec:scale-setting},
we fix the scale of lattice spacing $a$ and show our 
results of hadron spectra, 
%$f_\pi$, and $f_K$ 
decay constants of pseudoscalar mesons
and quark masses in physical units. 
Sec.~\ref{sec:summary} is devoted to summary and discussion.

%%%%%%%%%%%%%%%%%%%%%%%%%%%%%%%%%%%%%%
%%%%%%%%%%%%%%%%%%%%%%%%%%%%%%%%%%%%%%
\section{Simulation details}\label{sec:setup}
%%%%%%%%%%%%%%%%%%%%%%%%%%%%%%%%%%%%%%
%%%%%%%%%%%%%%%%%%%%%%%%%%%%%%%%%%%%%%
We generate $(2+1)$-flavor QCD configurations employing the Iwasaki gauge~\cite{Iwasaki:1983iya} and $\mathcal{O}(a)$-improved Wilson-clover quark actions. 
The lattice extent is $96^4$ and the lattice bare coupling constant $\beta\equiv6/g^2=1.82$ following Refs.~\cite{Ishikawa:2015rho, Ishikawa:2018jee, PACS:2019ofv}.

The quark action is given by
\beq
S_{q=u,d,s} &=& \sum_n \bar{q}_n \left[q_n -\kappa_q c_{\text{SW}} \frac{i}{2} \sum_{\mu, \nu} \sigma_{\mu \nu} F_{\mu \nu}(n) q_n \right.\nonumber\\
&& \left. -\kappa_q \sum_\mu \left\{ (1-\gamma_\mu) U_{n,\mu} q_{n+\hat{\mu}} + (1+\gamma_\mu) U^\dag_{n-\hat{\mu},\mu} q_{n-\hat{\mu}}  \right\} \right], \nonumber\\
\eeq
where the gauge field is $6$ times smeared using the stout smearing parameter $\rho=0.1$.
%where the field strength $F_{\mu\nu}$ given by clover terms.
We utilize $c_{\text{SW}}=1.11$, which is nonperturbatively determined by the Schr\"{o}dinger functional scheme in Ref.~\cite{Taniguchi:2012gew}.
The hopping parameters for $u,d$ quarks and $s$ quark are set to $(\kappa_{ud},\kappa_s)=(0.126117,0.124902)$, with which the hadron masses are reported to be almost the values 
at the physical point in Refs.~\cite{Ishikawa:2018jee, PACS:2019ofv}.
In our work,
configurations are generated through 5 independent Markov chains ($5$-run series)
by the hybrid Monte Carlo (HMC) algorithm,
where different random number seeds and different initial configurations
are used for each run.
As initial configurations, we pick 5 configurations from the K-conf,
which were generated with $(\kappa_{ud}, \kappa_s) = (0.126117, 0.124790)$
corresponding to slightly heavier quark masses than the physical point~\cite{Ishikawa:2015rho}.
After discarding more than $300$ trajectories for the thermalization process in each run, we generate $1,600$ trajectories in each run, thus, $8,000$ trajectories in total. We save the configurations every $5$ trajectories and use them for the measurement of physical quantities, %observables, 
e.g., topological charge and hadron correlation functions.

Let us explain how to implement the quark action in 
HMC in detail.
We symbolically rewrite the Wilson-clover operator as
\beq
D= 1+ T + M,
\eeq
where $T$ and $M$ denote the clover and the hopping terms, respectively.
As for the $u,d$ quarks with degenerated masses, we divide 
%the degenerated $2$ quark action, 
the corresponding quark action,
$| \det D |^2$, into several factors as follows.
First of all, factorization of the clover term as $D=(1+T)(1+(1+T)^{-1} M)\equiv (1+T)\tilde{D}$ gives
\beq
| \det D |^2 =
| \det (1+T) |^2 | \det \tilde{D} |^2.
\eeq
Since the term $(1+T)$ consists of the local quantities, it is easy to simulate 
%this term.
the first factor in the determinant.
On the other hand, the calculation of the second factor is expensive, 
so that we utilize the domain-decomposed HMC (DDHMC) algorithm~\cite{Luscher:2003vf}, 
the Hasenbusch mass preconditioning~\cite{Hasenbusch:2001ne, Hasenbusch:2002ai}, 
and the even-odd preconditioning.

The Dirac operator $\tilde{D}$ is decomposed by the even ($E$) and odd ($O$) domains,
\beq
\tilde{D}=\begin{pmatrix}
\tilde{D}_{EE} & \tilde{D}_{EO} \\
\tilde{D}_{OE} & \tilde{D}_{OO} \\
\end{pmatrix}.
\eeq
Utilizing the Schur decomposition, we rewrite  $| \det \tilde{D} |^2$  as 
\beq
\left| \det \tilde{D} \right|^2&& = 
\left| \det \tilde{D}_{EE} \right|^2 
\left| \det \tilde{D}_{OO} \right|^2 %\nonumber\\
%&& \times 
\left| \det [1-(\tilde{D}_{EE}^{-1} \tilde{D}_{EO} \tilde{D}_{OO}^{-1} \tilde{D}_{OE})] \right|^2.\nonumber
\eeq
Introducing the even-odd site decomposition in each domain and the spin projection that connects even-odd sites,
each determinant can be written by
\beq
&&\det \tilde{D}_{EE} = \det[(\hat{D}_{EE})_{ee}], \nn
&&\det \tilde{D}_{OO} = \det[ (\hat{D}_{OO})_{ee}],\nonumber\\
&&\det [1-(\tilde{D}_{EE}^{-1} \tilde{D}_{EO} \tilde{D}_{OO}^{-1} \tilde{D}_{OE})] = \det [ \hat{D}_{EE}^{\mathrm{spin}}],
\eeq
where we refer to Ref.~\cite{PACS-CS:2008bkb} 
for the explicit form of the spin projection.

Now, we can implement the $u,d$ quark action using three pseudo fermions ($\phi_{O_e}, \phi_{Ee}$ and $\chi_E$), 
\begin{widetext}
\beq
S_{ud} &=& - \log \left( | \det D |^2 \right) \nonumber\\
&\rightarrow& 
-2 \mathrm{Tr} \log (1+T) 
+ \sum_{X=E,O} \left| ((\hat{D}_{XX})_{ee})^{-1} \phi_{Xe} \right|^2 
+ \left| (\hat{D}_{EE}^{\mathrm{spin}})^{-1} \chi_E \right|^2 \nonumber\\
&\coloneqq& S_{ud, \mathrm{clv.}}[U; \kappa_{ud}] + \sum_{X=E,O} S_{ud,\mathrm{UV}} [U,\phi_{Xe}; \kappa_{ud}] + S_{ud,\mathrm{IR}}[U,\chi_{E}; \kappa_{ud}].
\eeq
\end{widetext}
The calculations of the first and second terms can be performed within each domain, while the third term describes the long-range modes and 
the simulation of this term requires data transfer between different domains.
Therefore, we introduce the two-fold Hasenbusch preconditioning to speed up the simulation.
By introducing $\kappa_1= \rho_1 \kappa_{ud}$ and $\kappa_2 = \rho_2 \kappa_1$
($\rho_1,\rho_2 <1$)
with two additional pseudo fermions, $\zeta_E^1, \zeta_E^2$,
the third term is decomposed into three parts,
\beq 
\left| (\hat{D}_{EE}^{\mathrm{spin}} (\kappa_{ud}))^{-1} \chi_E\right|^2
&\rightarrow& 
\left|\frac{\hat{D}_{EE}^{\mathrm{spin}} (\kappa_1) }{\hat{D}_{EE}^{\mathrm{spin}} (\kappa_{ud})} \zeta_E^1 \right|^2 %\nonumber\\
%&&
+ \left|\frac{\hat{D}_{EE}^{\mathrm{spin}} (\kappa_2) }{\hat{D}_{EE}^{\mathrm{spin}} (\kappa_1)} \zeta_E^2 \right|^2 
+ \left| (\hat{D}_{EE}^{\mathrm{spin}} (\kappa_2) )^{-1} \chi_E \right|^2\nonumber\\
&\coloneqq& 
S_{{ud},\mathrm{IR1}} + S_{{ud},\mathrm{IR2}}+ S_{{ud},\mathrm{IR3}} 
\eeq
In this work, we tune the values of $\rho_1$ and $\rho_2$ as $\rho_1=0.99975$ and $\rho_2=0.99600$, respectively.

The strange quark part is implemented by the rational hybrid Monte Carlo (RHMC) algorithm~\cite{Clark:2006fx}.
The clover term is factored out as in the case of the $u,d$ quarks,
$D = (1+T) \tilde{D}$,
and the weight factor associated with the second term is evaluated with the rational approximation.
For the calculations of the force,
the range of the approximation
for 
the rational functions of 
%$1/\sqrt{x}$ with 
$x = \tilde{D}^\dag \tilde{D}$ is 
taken as $[x_{\rm min}, x_{\rm max}]=[0.000200000001, 2]$ with the order of $N_{\mathrm{\mathrm{RHMC,F}}}=10$ expansion,
which ensures $10^{-7}$ accuracy for $1/\sqrt{x}$.
Note that these parameters for the rational approximation of the force 
are fixed during the steps of molecular dynamics (MD)
in order to fix the equation of motion of the MD.
We also checked that our simulation is actually within the range of 
$[x_{\rm min}, x_{\rm max}]$.
On the other hand, 
in the calculation of the strange quark Hamiltonian for the Metropolis test, the approximation range is
chosen from min and max eigenvalues of the $\tilde{D}^\dag \tilde{D}$ operator
(with an additional margin factor of $1.01$ for both min and max),
%extended by the factor of $1.01$, 
and the order of the approximation, $N_{\mathrm{RHMC,H}}$,
is tuned to keep $10^{-14}$ accuracy (typically $N_{\mathrm{RHMC,H}}=20$ is chosen.).

\begin{figure}[htbp]
    \begin{tabular}{c}
        \includegraphics[keepaspectratio, scale=0.7]{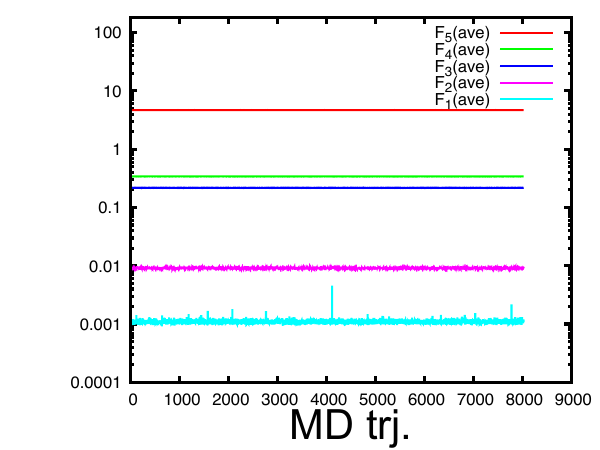}
    \end{tabular}
            \caption{History of the force average in the MD evolution.}\label{fig:hist-force-ave}
 \end{figure}

\begin{figure}[htbp]
    \begin{tabular}{c}
        \includegraphics[keepaspectratio, scale=0.3]{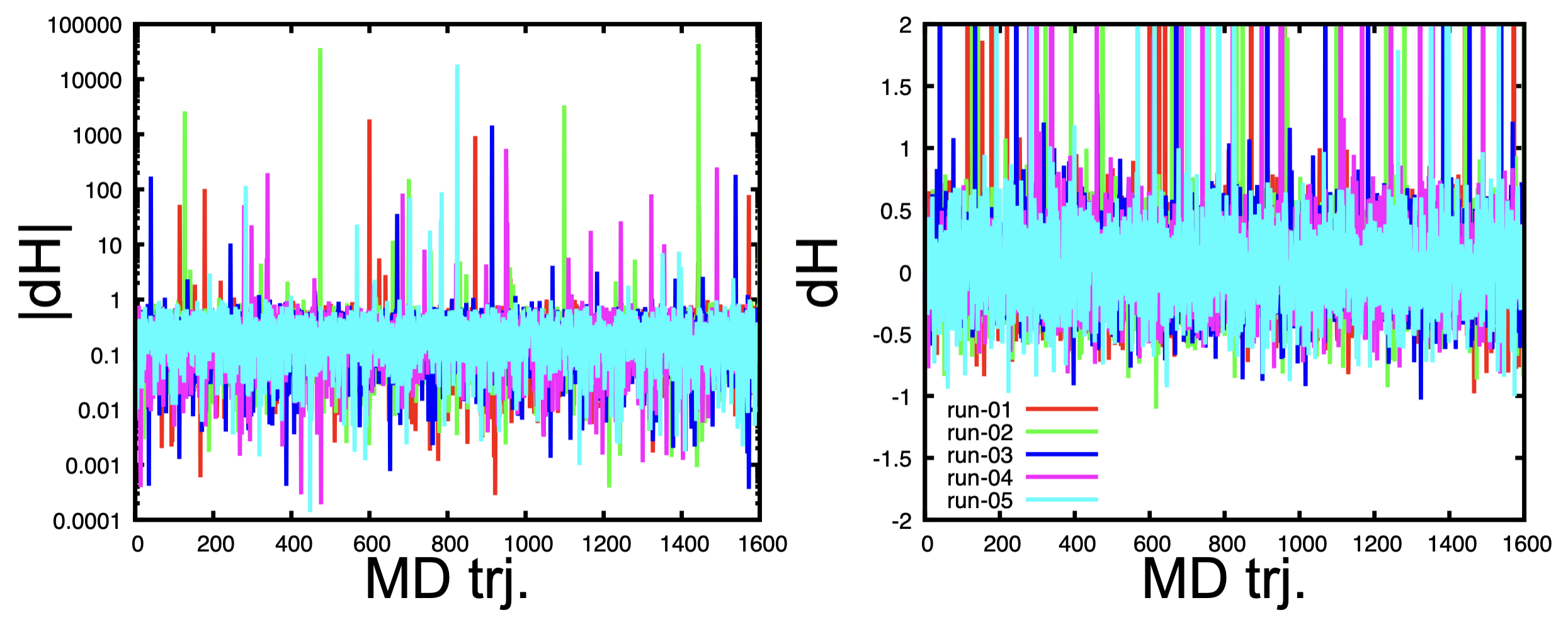}
    \end{tabular}
            \caption{History of difference of Hamiltonian, $dH$, in the MD evolution for each run,
            shown by
            a logarithmic plot of $|dH|$ (left)
            and an enlarged plot of $dH$ around $dH = 0$ (right).
            }
            \label{fig:hist-dH}
 \end{figure}

Corresponding to each term in the action,
we have the following force terms in the MD steps in the HMC algorithm:
the gauge force $F_g$ from the gauge action, 
the clover force $F_{ud, \mathrm{clv.}}$, 
the UV force $F_{ud,\mathrm{UV}}$ and 
the three IR parts 
$F_{ud,\mathrm{IR1}}$, $F_{ud,\mathrm{IR2}}$, $F_{ud,\mathrm{IR3}}$
from the $u,d$ quark actions,
and
the clover force $F_{s,\mathrm{clv.}}$ and
the RHMC force $F_{s,\mathrm{RHMC}}$ from the $s$ quark action.
In the MD steps,
we adopt the multiple time scale integration scheme~\cite{Sexton:1992nu}
with a depth of 5,
where the hierarchy of forces is taken as follows:
$F_5 = F_g$,
$F_4 = F_{ud,\mathrm{UV}}$,
$F_3 = F_{ud,\mathrm{IR3}} + F_{s,\mathrm{RHMC}} + F_{ud, \mathrm{clv.}} + F_{s,\mathrm{clv.}}$
$F_2 = F_{ud,\mathrm{IR2}}$,
$F_1 = F_{ud,\mathrm{IR1}}$,

We employ the leapfrog algorithm for the MD evolution,
as in the case of Refs.~\cite{PACS-CS:2008bkb, PACS-CS:2009sof, Ishikawa:2015rho, Ishikawa:2018jee, PACS:2019ofv}
\footnote{N. Ukita, private communications.},
since we find that the Omelyan algorithm is less stable.
In the solver for the pseudo fermions, we impose the stopping condition 
of $|r_{\rm crit}| < 10^{-14}$ for the evaluations of both the forces and the Hamiltonian
so that the reversibility in the MD evolution 
and the accuracy of the Metropolis test
are assured~\cite{PACS-CS:2008bkb, PACS-CS:2009sof, Ishikawa:2015rho}.
For the IR part of the $u,d$ quarks,
we employ the mixed precision nested BiCGStab solver
with the Schwarz Alternating Procedure (SAP)
and point Jacobi iteration preconditioners~\cite{PACS-CS:2008bkb, Ishikawa:2021iqw}. 
The code optimization for Fugaku developed in the QWS library~\cite{Ishikawa:2021iqw} is utilized in the quark solver.
Within one MD trajectory ($\delta \tau_{\mathrm{MD}}=1.0$), 
the time step of the MD evolution is implemented using a set of integers ($N_5,N_4,N_3,N_2,N_1$) for each force calculation:
$\delta \tau_5 = \delta \tau_{\mathrm{MD}}/(N_5 N_4 N_3 N_2 N_1)$, 
$\delta \tau_4 = \delta \tau_{\mathrm{MD}}/(N_4 N_3 N_2 N_1)$, 
$\delta \tau_3 = \delta \tau_{\mathrm{MD}}/(N_3 N_2 N_1)$,
$\delta \tau_2 = \delta \tau_{\mathrm{MD}}/(N_2 N_1)$, and 
$\delta \tau_1 = \delta \tau_{\mathrm{MD}}/N_1$.
We take ($N_5,N_4,N_3,N_2,N_1$) $=$ ($8,2,2,2,22$).
Figure~{\ref{fig:hist-force-ave}} depicts the history of the force
average in each MD trajectory. 
%of our calculations. 
%
Figure~\ref{fig:hist-dH} represents the difference of Hamiltonian, $dH$, for each MD trajectory. 
While there are several spikes in total $8,000$ trajectories, 
the simulation is basically stable during the MD evolution.
It results in the $87$\% acceptance rate.

%%%%%%%%%%%%%%%%%%%%%%%%%%%%%%%
 %%%%%%%%%%%%%%%%%%%%%%%%%%%%%%%%%%%%%

%%%%%%%%%%%%%%%%%%%%%%%%%%%%%%%
 %%%%%%%%%%%%%%%%%%%%%%%%%%%%%%%%%%%%%

\begin{figure}[htbp]
    \begin{tabular}{c}
        \includegraphics[keepaspectratio, scale=0.75]{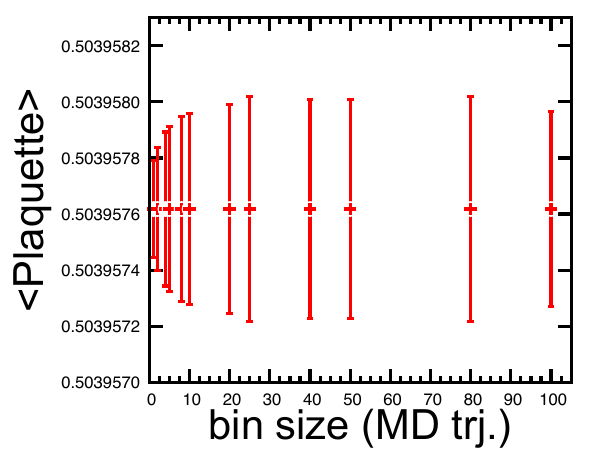}
    \end{tabular}
            \caption{Bin-size dependence in the jackknife analysis for the plaquette average.}\label{fig:hist-plaq}
 \end{figure}

In Fig.~\ref{fig:hist-plaq}, we show the plaquette values
calculated by the jackknife method with different bin sizes.
As a central value, we take the bin size $=100$ trajectories and obtain $\langle \mathrm{plaquette} \rangle = 0.5039576(3)$.
Our result is $1 \sigma$ consistent with the data by the PACS Collaboration on $64^4$ and $128^4$ lattices with the same $\beta$ and $\kappa_{ud},\kappa_s$~\cite{Ishikawa:2018jee, PACS:2019ofv}, 
but the statistical errors of our data are much smaller thanks to 
our higher statistics.
%For a precise estimation of the auto correlation, we will study an integrated auto-correlation time in the next section.
%
We summarize basic data for the configuration generation in Table~\ref{table:simulation-data}.
All statistical errors are estimated by the jackknife method with 
the bin size of $100$ trajectories.

%%%%%%%%%%%%%%%%%%%%%%%%%%%%%%%
 %%%%%%%%%%%%%%%%%%%%%%%%%%%%%%%%%%%%%

%%%%%%%%%%%%%%%%%%%%%%%%%%%%%%%%%%%%%%
%%%%%%%%%%%%%%%%%%%%%%%%%%%%%%%%%%%%%%
%%%%%%%%%%%%%%%%%%%%%%%%%%%%%%%%%%%%%%
\begin{widetext}
\begin{table*}[t]
\centering
\begin{tabular}{cccccccc}
  \hline \hline 
$\langle$ plaquette $\rangle$ & $\langle e^{-dH} \rangle$ & acc. rate &  $\langle F_5 \rangle$  & $\langle F_4 \rangle$  & $\langle F_3 \rangle$ & $\langle F_2 \rangle$ & $\langle F_1 \rangle$  \\
0.5039576(3) & 0.991(4) & 0.872 &4.689783(1) & 0.340328(1) &0.217155(2)  & 0.009101(5) & 0.001096(1)\\
 \hline \hline 
\end{tabular}
\caption{ Detail data of configuration generation } \label{table:simulation-data}
\end{table*}
\end{widetext}
%%%%%%%%%%%%%%%%%%%%%%%%%%%%%%%%%%%%%%
%%%%%%%%%%%%%%%%%%%%%%%%%%%%%%%%%%%%%%
%%%%%%%%%%%%%%%%%%%%%%%%%%%%%%%%%%%%%%

%%%%%%%%%%%%%%%%%%%%%%%%%%%%%%%%%%%%%%
%%%%%%%%%%%%%%%%%%%%%%%%%%%%%%%%%%%%%%
\section{Ensemble properties: auto correlation}\label{sec:property}
%%%%%%%%%%%%%%%%%%%%%%%%%%%%%%%%%%%%%%
%%%%%%%%%%%%%%%%%%%%%%%%%%%%%%%%%%%%%%
In this section, we show the ensemble properties of generated configurations by investigating 
several local observables such as the energy density as a function of the gradient flow time
and the topological charge distribution.
Furthermore, we study the integrated auto-correlation time of these quantities.
As will be shown below, the resulting integrated auto-correlation times 
are found to be reasonably short (less than $10$ trajectories)
even for the topological charges. 
This indicates that an independent configuration set has been generated.

To measure the topological charge distribution, we adopt the definition of the topological charge through the gradient flow~\cite{Luscher:2010iy}.
The gradient flow is defined by the following equations
\beq
\partial_t V_t (x,\mu) &=& - g_0^2 \{ \partial_{x,\mu} S (V_t) \} V_t (x,\mu),\label{eq:Wflow}\\
V_t(x,\mu)|_{t=0} &=& U(x,\mu), 
\eeq
where $x = (\vec{x}, \tau)$ is the space-time lattice point and $t$ is the flow-time, $V_t(x,\mu)$ denotes the smeared link variables over a radius $r=\sqrt{8t}$ at the flow-time $t$, and $U(x,\mu)$ is the generated link variable in Monte Carlo calculations.
Here, $S$ represents the flow gauge action, which we can choose independently from the gauge action in the configuration generation process.
In our work, we utilize the standard plaquette gauge action as a flow gauge action to solve the differential equation Eq.~\eqref{eq:Wflow}.
We use the third-order Runge-Kutta algorithm with a step size  $\varepsilon = 0.01$.

The gluonic definition of the topological charge density can be given by
\beq
q(x,t) = -\frac{1}{32 \pi^2} \epsilon_{\mu \nu \rho \sigma} \mathrm{tr} \{ G_{\mu \nu} (x,t) G_{\rho \sigma}(x,t) \},\label{eq:topo-density}
\eeq
with a clover-type definition of the field strength $G_{\mu \nu}(x,t)$, which is constructed by the smeared link variables $V_t$.
The topological charge can be directly calculated by the topological charge density in Eq.~\eqref{eq:topo-density},
\beq
Q(t)= a^4 \sum_x q(x,t).
\eeq
The value of $Q(t)$ roughly plateaus in a long $t$ region, but small fluctuations exist. Therefore, we introduce a reference scale $t_0$ and identify the value of $Q(t=t_0)$ as a convergent value of $Q$ for each configuration~\cite{Bruno:2014ova}.

The reference scale $t_0$ is originally introduced in Ref.~\cite{Luscher:2010iy}, which is given by
\beq
t^2 \langle E(t) \rangle |_{t=t_0} =0.3,
\eeq
where $E(t)$ denotes the energy density
\beq
E(t)= -\frac{1}{2V} \sum_x \mathrm{tr} \{ G_{\mu \nu} (x,t) G_{\mu \nu}(x,t) \}.
\eeq
Here, we utilize a clover-type definition for $G_{\mu \nu} (x,t)$ again. 
%%%%%%%%%%%%%%%%%%%%%%%%%%%%%%%
 \begin{figure}[t]
 \centering
    \begin{tabular}{c}
        \includegraphics[keepaspectratio, scale=0.75]{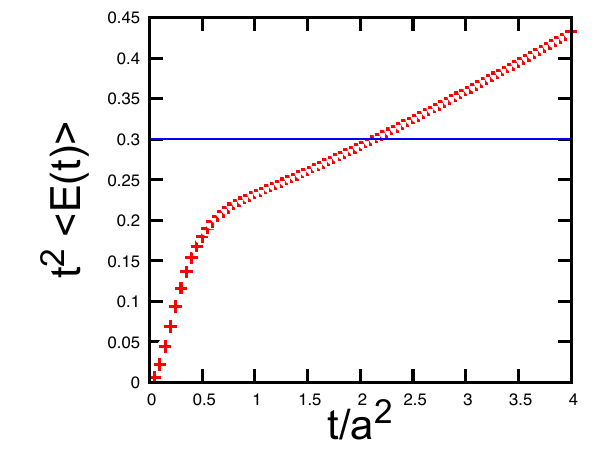}
    \end{tabular}
            \caption{Data of $t^2 \langle E(t) \rangle$ as a function of the gradient flow time, $t/a^2$. }\label{fig:t2E}
 \end{figure}
 %%%%%%%%%%%%%%%%%%%%%%%%%%%%%%%%%%%%%
 Figure~\ref{fig:t2E} depicts the dimensionless energy density $t^2 \langle E(t) \rangle$ as a function of the flow-time $t$.
 We obtain  
 \beq
 t_0/a^2 = 2.1047(4)
 \eeq
 as an averaged value of $t_0$ in lattice unit results. 
We also calculate a similar reference scale,
\beq
w_0/a = 2.0126(4),
\eeq
which is defined by~\cite{BMW:2012hcm}
\beq
\left. t \frac{d}{dt}  \left\{ t^2 \langle E(t) \rangle \right\} \right|_{t=w_0^2} =0.3.
\eeq

\begin{figure}[htbp]
    \begin{tabular}{c}
        \includegraphics[keepaspectratio, scale=0.75]{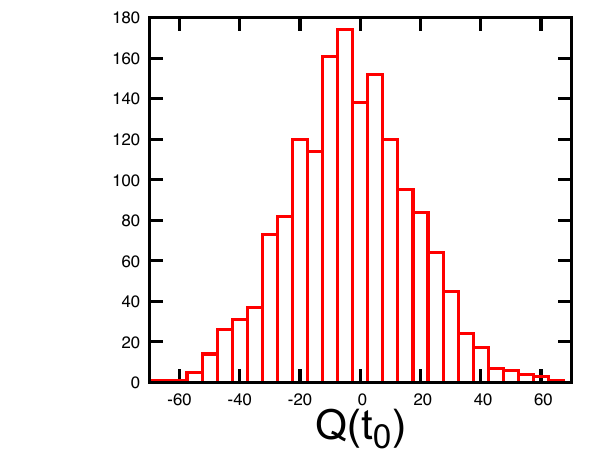}
    \end{tabular}
            \caption{Histogram of the topological charge $Q(t)$ at $t=t_0$.}
            \label{fig:histogram-Q}
 \end{figure}

Figure~\ref{fig:histogram-Q} represents the histogram of $Q(t_0)$ for total $1,600$ generated configurations.
 %%%%%%%%%%%%%%%%%%%%%%%%%%%%%%%
 %%%%%%%%%%%%%%%%%%%%%%%%%%%%%%%%%%%%%
It shows almost a Gaussian distribution as expected.
Table~\ref{table:topo} is the summary of the averaged values of $Q(t_0)$ and $Q(t_0)^2$, and the susceptibility $\chi_Q= \langle (Q-\langle Q \rangle)^2/V$.
As expected, $\langle Q \rangle $ is consistent with zero within $1.1 \sigma$ uncertainty.
%%%%%%%%%%%%%%%%%%%%%%%%%%%%%%%%%%%%%%
%%%%%%%%%%%%%%%%%%%%%%%%%%%%%%%%%%%%%%
%\begin{widetext}
\begin{table*}[t]
%\centering
\begin{tabular}{ccc}
  \hline \hline 
$\langle Q \rangle$ & $\langle Q^2 \rangle$ & $\chi_Q$  \\
$-0.6644(6167) \quad$ & $ \quad416.8(16.1) \quad$ &  $ \quad4.90 (19)  \times 10^{-6} $\\
 \hline \hline 
\end{tabular}
\caption{ Averages and statistical errors of $Q$, $Q^2$ and the topological susceptibility $\chi_Q$.} \label{table:topo}
\end{table*}
%\end{widetext}
%%%%%%%%%%%%%%%%%%%%%%%%%%%%%%%%%%%%%%
%%%%%%%%%%%%%%%%%%%%%%%%%%%%%%%%%%%%%%

Now, let us estimate the auto-correlation times of local observables.
We follow the papers~\cite{RBC:2012cbl, RBC:2014ntl} to define the integrated auto-correlation time and its error estimation.
The integrated auto-correlation time $\tau_{\mathrm int.}$ for the observable $A$ 
as a function of the cutoff in MD time separation ($\Delta_{\mathrm{cut}}$)
is given by
\beq
\tau_{\mathrm{int.}}(\Delta_{\mathrm{cut}})=\frac{1}{2}+\sum_{\Delta = 1}^{\Delta_{{\mathrm{cut}}}} C(\Delta),
\eeq
where 
\beq
C(\Delta)=\left\langle \frac{(A_i -\bar{A}) (A_{i+\Delta} - \bar{A})}{\sigma^2}  \right\rangle_i 
\label{eq:def-C-delta}
\eeq
with the index $i$ denoting the MD time.
Here, $\bar{A}$ and $\sigma$ represent the mean and variance of $A_i$. We first calculate them using the jackknife method with a sufficiently large bin size. By referring to
Fig.~\ref{fig:hist-plaq}, we take $20$ MD trajectories in our analyses. Next, we bin the data inside the brackets in Eq.~\eqref{eq:def-C-delta} for a fixed $\Delta$. Here, the bin size can be taken independently of the first bin size in the jackknife method. We increase it until the statistical error of $\tau_{\mathrm{int.}}$ saturates.
The statistical error of $\tau_{\mathrm{int.}}$  is estimated by the bootstrap method by resampling the binned data inside the brackets in Eq.~\eqref{eq:def-C-delta}.

%%%%%%%%%%%%%%%%%%%%%%%%%%%%%%%
%\begin{widetext}
 \begin{figure}[htbp]
 \begin{center}
        \includegraphics[keepaspectratio, scale=0.35]{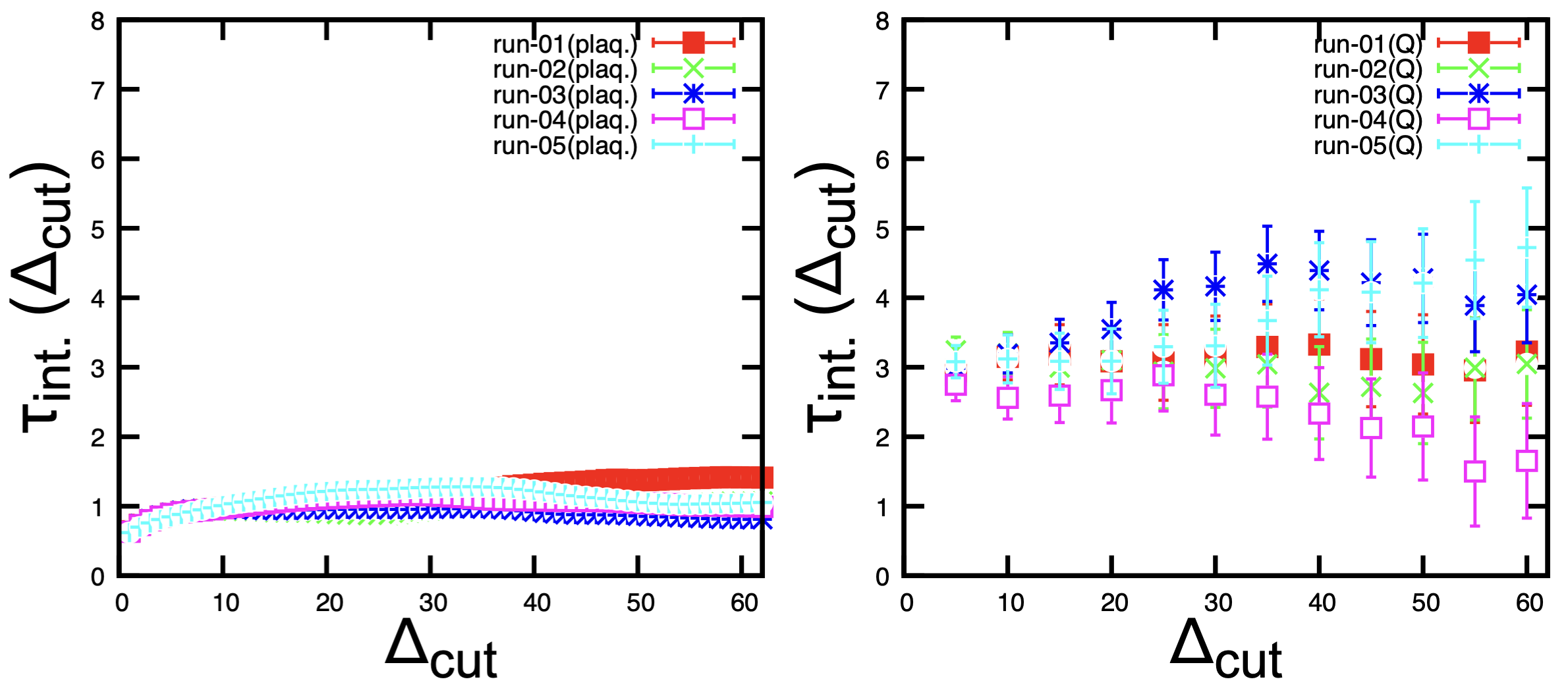}
    \caption{
Integrated auto-correlation time $\tau_{\mathrm{int.}}$ for the plaquette (left) and topological charge ($Q(t_0)$) (right) as a function of $\Delta_{\mathrm{cut}}$
for each run.
}\label{fig:AC-time}
\end{center}
  \end{figure}
%\end{widetext}
 %%%%%%%%%%%%%%%%%%%%%%%%%%%%%%%%%%%%%
Figure~\ref{fig:AC-time} displays the integrated auto-correlation times for the plaquette and topological charge ($Q(t_0)$) in the left and right panels, respectively.
Here, we analyze our data for each run series independently. We observe that the obtained integrated auto-correlation times saturate in short trajectory regions, namely at most $2$ MD trajectories for the plaquette and $6$ MD trajectories for the topological charge. 

\begin{figure}[htbp]
    \begin{tabular}{c}
        \includegraphics[keepaspectratio, scale=0.75]{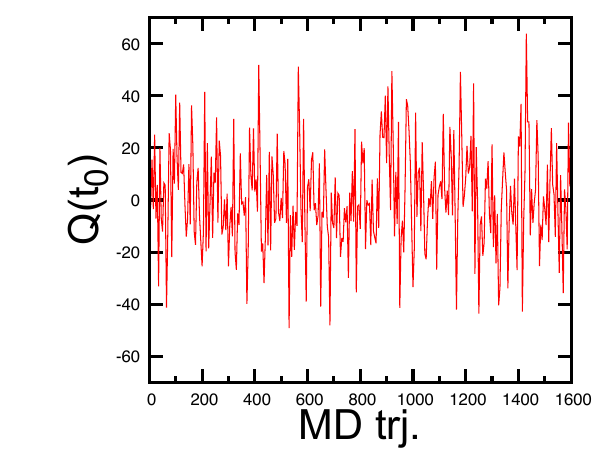}
    \end{tabular}
            \caption{History of the topological charge $Q(t_0)$ in the MD evolution for the run-01 configurations.}\label{fig:raw-data-Q}
 \end{figure}

In order to further confirm such a short auto-correlation time,
we study the history of $Q$ in the MD evolution.
As shown in Fig.~\ref{fig:raw-data-Q}, we find that
the value of $Q$ indeed changes significantly even for $1$ configuration separation
corresponding to 5 MD trajectory separation.

%%%%%%%%%%%%%%%%%%%%%%%%%%%%%%%
%%%%%%%%%%%%%%%%%%%%%%%%%%%%%%%%%%%%%

There are two possible reasons for this short auto-correlation time.
One is the use of the Wilson fermions, where the near-zero mode is absent even in a small quark mass regime. 
The other is that the lattice spacing in our simulation,
which will be determined as $a\approx 0.08$fm later in this paper,
is relatively coarse compared to the spacing where the topological freezing starts to appear.
Consequently, the short auto-correlation time suggests that our set of $1,600$ configurations is well spread in the configuration space
and serves as good ensembles for lattice QCD measurements of physical quantities.

%%%%%%%%%%%%%%%%%%%%%%%%%%%%%%%%%%%%%%
%%%%%%%%%%%%%%%%%%%%%%%%%%%%%%%%%%%%%%
\section{Stable Mesons}
\label{sec:meson-corr}

%%%%%%%%%%%%%%%%%%%%%%%%%%%%%%%%%%%%%%
%%%%%%%%%%%%%%%%%%%%%%%%%%%%%%%%%%%%%%
\subsection{Calculation strategy}
\label{subsec:meson-strategy}
We study the meson spectra, the decay constants and the PCAC masses for $u,d$ and $s$ quarks from the two-point correlation functions.

The local operators for flavor off-diagonal pseudo-scalar (PS) and axial-vector current 
with zero-momentum projection
are defined by
\beq
P(\tau) &=& \sum_{\vec{x}} \bar{q}_f (\vec{x},\tau) \gamma_5 q_g(\vec{x},\tau), \label{eq:def-PS-meson} \\
A_\mu (\tau) &=& \sum_{\vec{x}} \bar{q}_f (\vec{x},\tau) \gamma_\mu \gamma_5 q_g (\vec{x},\tau).\label{eq:def-A4}
\eeq
Here, the indices $f,g$ with $f\neq g$ label the flavor of the valence quarks ($f,g= u, d, s$).
In terms of 
the operator improvement of the axial-vector current,
$A_\mu (\tau) \rightarrow A_\mu(\tau) + a c_A \partial_\mu P(\tau)$,
we employ the tree-level value, $c_A = 0$,
since it is expected to be a good approximation in our lattice setup
thanks to the link smearing 
for the fermion~\cite{BMW:2010skj, BMW:2013fzj},
and the explicit non-perturbative calculation
found that $c_A$ is indeed consistent with zero 
within the statistical error~\cite{Taniguchi:2012gew}.

To measure the correlation function between the currents, we employ the wall source method without gauge fixing~\cite{Kuramashi:1993ka}.
Quark propagators are solved 
with the periodic boundary condition for all directions,
where
the stopping conditions are taken as
$|\rcrit| < 10^{-6}$ for $u,d$ quarks and 
$|\rcrit| < 10^{-12}$ for $s$ quark. 
%$\rcrit = |D\vec{x} - \vec{b}|/|\vec{b}|$
We have checked that these stopping conditions are sufficiently precise 
in our study.
To suppress statistical fluctuations, we take the average between forward and backward propagation, and we also measure the correlation function in each of the four directions for one configuration using the fact that our lattice has hypercubic symmetry. 
Thus, the total number of measurements for the observables in this section is 
$1600 \times 4 \times 2 = 12,800$.
The statistical errors in this section 
are estimated by the jackknife method.

To obtain the PS meson masses ($m_{\mathrm{PS}}$), the PCAC quark masses
and the PS meson decay constants, we utilize 
the PS-PS correlators and PS-A$_4$ correlators as follows.
The PS-PS correlators 
at $\tau/a, (T-\tau)/a \rightarrow \infty$
are expressed by
%\begin{align}
\beq
\langle P(\tau) P^\dag(0) \rangle &\simeq& N_s^3 C_{PP}W(m_{\mathrm{PS}}T) %\nonumber\\
%& 
\times 
\left[ \exp(-m_{\mathrm{PS} }\tau) +\exp(-m_{\mathrm{PS}} (T-\tau)) \right],  %\nonumber\\
\label{eq:PS-PS}
\eeq
%\end{align}
where $W(m_{\mathrm{PS}}T)$ with $T/a=96$ counts the contribution from wrapping propagation of the PS meson around the lattice in the temporal direction. This is given as the sum of zero to infinite windings,
%\begin{align}
\beq
W(m_{\mathrm{PS}}T) &=& 1+\exp(-m_{\mathrm{PS}}T) + \exp(-2m_{\mathrm{PS}}T) + \cdots, \nonumber\\
&=& \frac{1}{1-\exp(-m_{\mathrm{PS}}T)}.
\eeq
%\end{align}
Similarly, the PS-A$_4$ correlators 
at $\tau/a, (T-\tau)/a \rightarrow \infty$
have the following form,
%\begin{align}
\beq
\langle A_4 (\tau) P^\dag (0) \rangle &\simeq& N_s^3 C_{AP} W(m_{\mathrm{PS}} T ) %\nonumber\\
%&= 
\times 
\left[\exp (-m_{\mathrm{PS}} \tau) - \exp(-m_{\mathrm{PS}}(T - \tau))\right]. %\nonumber\\ 
\label{eq:PS-A4}    
\eeq
%\end{align}

We define the PCAC bare quark masses 
through axial Ward identity (AWI)
as
\beq
m_{f}^{\rm AWI}+m_g^{\rm AWI} = \frac{\langle 0| \partial_4 A_4 | PS \rangle  }{\langle 0 | P | PS \rangle }.\label{eq:def-mq-local}
\eeq
They can be evaluated with 
\beq
m_f^{\rm AWI} + m_g^{\rm AWI} = m_{\mathrm{PS}} \left| \frac{C_{AP}}{C_{PP}} \right|,
\label{eq:def-mq}
\eeq
using the parameters given in Eqs.~\eqref{eq:PS-PS} and \eqref{eq:PS-A4}.

As for the decay constants of the PS mesons, we utilize the following expression
for the bare values,
\beq
f_{\mathrm{PS}, fg}^{\rm bare}= 
\sqrt{2\kappa_f}\sqrt{2\kappa_g} 
\frac{\sqrt{2} |C_{AP}|}{\sqrt{m_{\mathrm{PS}} |C_{PP}|}} .
\quad
\label{eq:def-fPS}
\eeq

In our analysis, we first study the effective masses of the PS mesons,
defined locally in time using two data points, 
$\tau$ and $\tau + a$ of Eq.~\eqref{eq:PS-PS} (or Eq.~\eqref{eq:PS-A4}),
and examine the ground state saturation.
The effective PCAC quark masses are studied 
by using Eq.~\eqref{eq:def-mq-local}, where the derivative of $A_4$ is estimated by the 
forth-order symmetric difference.

To obtain our main results, 
we perform simultaneous fit of Eqs.~\eqref{eq:PS-PS} and \eqref{eq:PS-A4}
to determine best-fit values of $m_{\mathrm{PS}}, C_{PP}$ and $C_{AP}$,
from which the PS meson masses and the bare values of the PCAC masses and decay constants are obtained.

In order to obtain the physical values of the quark masses and decay constants,
we consider the ${\cal O}(ma)$ improvement and the renormalization factors.
The expressions for the ${\cal O}(ma)$ improvement are given by~\cite{Bhattacharya:2005rb} 
\beq
m_{fg}^{\rm imp.} = 
\left[
\frac
{1 + a b_A m_{fg} + a \bar{b}_A {\rm Tr}M}
{1 + a b_P m_{fg} + a \bar{b}_P {\rm Tr}M}
\right] 
m_{fg}^{\rm AWI},
\label{eq:mq_O(ma)}
\eeq
\beq
f_{\mathrm{PS}, fg}^{\rm imp.} =
\left[ 1+ a b_A m_{fg} + a \bar{b}_A {\rm Tr}M \right] f_{\mathrm{PS}, fg}^{\rm bare} ,
\label{eq:f_O(ma)}
\eeq
where
$m_{fg} \equiv (m_f + m_g)/2$ and
$M \equiv {\rm diag}(m_u, m_d, m_s)$ represent the bare quark masses, and
$b_A, \bar{b}_A$ and $b_P, \bar{b}_P$ are improvement coefficients
for axial-vector and pseudo-scalar currents, respectively.
Note that these formulae are valid for flavor non-singlet operators~
\footnote{
In the literature, a similar ${\cal O}(ma)$-improvement formula 
for the individual quark mass is sometimes employed~\cite{PACS-CS:2008bkb, BMW:2010skj}.
However, such a formula cannot be justified,
since individual quark mass contains both flavor singlet and non-singlet
component.
}.

As regards the bare quark masses in the improvement terms, $m_f$, one may employ
those associated with axial Ward identity (AWI), $m_f^{\rm AWI}$, or
with vector Ward identity (VWI), 
$\displaystyle m_f^{\rm VWI} \equiv \frac{1}{a}\left(\frac{1}{2\kappa_f} - \frac{1}{2\kappa_c}\right)$
with $\kappa_c$ being the critical hopping parameter, 
as far as the improvement coefficients are defined consistently.
The relation between the two definitions is given as~\cite{Bhattacharya:2005rb}
\beq
m_{fg}^{\rm AWI} &=& \frac{Z_P Z_m}{Z_A} \left[ m_{fg}^{\rm VWI} 
+ (r_m - 1) \frac{{\rm Tr}M^{\rm VWI}}{N_f} \right] + {\cal O}(a), \\
{\rm Tr}M^{\rm AWI} &=& \frac{Z_P Z_m}{Z_A} r_m {\rm Tr}M^{\rm VWI} + {\cal O}(a) ,
\eeq
where
$Z_A, Z_P$ and $Z_m$ denote the renormalization factor 
in flavor non-singlet sector
for the axial-vector current, % operator, 
for the pseudo-scalar current, and for the VWI quark mass, respectively,
and
$r_m$ is the ratio of renormalization factors
between flavor singlet and non-singlet sectors
for the VWI quark mass.
In order to  evaluate the ${\cal O}(ma)$ improvement terms
in Eqs.~(\ref{eq:mq_O(ma)}) and (\ref{eq:f_O(ma)}),
we employ the tree-level values,
$Z_A = Z_P = Z_m = r_m = 1$
and $b_A = b_p = 1, \bar{b}_A = \bar{b}_P = 0$,
and
with 
$m_{f}$ substituted by $m_{f}^{\rm AWI}$.
In perturbation theory, the correction to the tree-level value 
is ${\cal O}(g_0^4)$ for $r_m, \bar{b}_A, \bar{b}_P$ 
and ${\cal O}(g_0^2)$ for others.
In our lattice setup, we expect 
such corrections are well suppressed thanks to the stout smearing.
In fact, in the case of $Z_A$ and $Z_P$,
the values determined non-perturbatively
are found to agree with the tree-level values
within a few percent errors~\cite{Ishikawa:2015fzw}.
Considering also that the magnitude of the improvement term is small,
we neglect the uncertainties associated with the tree-level evaluation of 
the ${\cal O}(ma)$ improvement.

Finally, we obtain the physical values for the quark masses and decay constants by
\beq
m_{fg}^{\rm \overline{MS}} &=& 
\frac{Z_A}{Z_P} m_{fg}^{\rm imp.}, \\
f_{\mathrm{PS}, fg} &=& Z_A f_{\mathrm{PS}, fg}^{\rm imp.}.
\eeq

In our calculation, we take 
$
(Z_A/Z_P)^{\rm \overline{MS}}
= Z_m^{\rm \overline{MS} \leftarrow SF} \cdot (Z_A/Z_P)^{\rm SF}
= 0.9950(142)
$
%(@@ = 0.9950(111)(89))
at $\mu=$ 2 GeV
and 
$Z_A=0.9650(117)$
%(@@ = 0.9650(68)(95) @@)$,
where
$Z_A, Z_P$ are determined nonperturbatively
in the Schr{\"o}dinger functional (SF) scheme
and 
$Z_m^{\rm \overline{MS} \leftarrow SF}$
is the mass matching factor between SF and 
${\rm \overline{MS}}$ schemes~\cite{Ishikawa:2015fzw}.
The uncertainties of the renormalization factors
are included in the systematic errors of our results.

In terms of the quark masses, 
we remark that there exists another formula called 
the ratio-difference method~\cite{BMW:2010skj}.
%, in which AWI and VWI bare quark masses are combined in a different way.
While we do not use this method in this work,
we give a quick summary for this method in Appendix~\ref{sec:ratio-diff},
where we argue that it is necessary to modify the formula known in the literature
in order to respect that Eq.~\eqref{eq:mq_O(ma)} is valid only for flavor non-singlet sectors.

\subsection{Numerical results: PS meson spectra and PCAC quark masses}
\label{subsec:PS-PCAC-masses}

\begin{figure}[h]
 \begin{center}
        \includegraphics[keepaspectratio, scale=0.7]{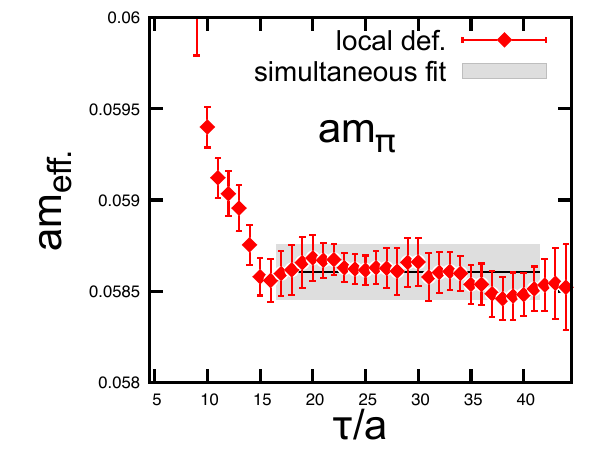}
        \includegraphics[keepaspectratio, scale=0.7]{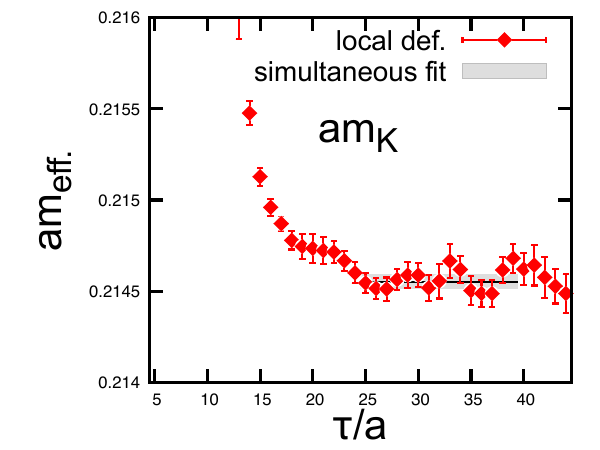}
    \caption{Effective masses for pion (left) and kaon (right) obtained from the PS-PS correlators.
    The black lines with gray bands depict the results with statistical errors
    obtained by the simultaneous fit 
    of the PS-PS and PS-A$_4$ correlators.
}\label{fig:effective-mass-PS}
\end{center}
  \end{figure}

Here, we present our numerical results. 
In Fig.~\ref{fig:effective-mass-PS},
we show the effective masses of the PS mesons
using the PS-PS correlators 
in the light-light quark sector 
($q_f, q_g = q_{u,d}$ in Eq.~(\ref{eq:PS-PS})) 
for pion 
and in the light-heavy quark sector
($q_f = q_{u,d}$ and $q_g = q_s$)
for kaon. 
We observe clear plateaux for both PS mesons. 
The results from the PS-A$_4$ correlators show similar behavior.

The PS meson masses are obtained by 
the simultaneous fit of the PS-PS and PS-A$_4$ correlators of Eqs.~\eqref{eq:PS-PS} and \eqref{eq:PS-A4},
where
the fit range is taken to be
$17 \le \tau/a \le 41$ for the light-light quark sector and
$25 \le \tau/a \le 39$ for the light-heavy quark sector.
The corresponding fit results as well as the statistical errors
are shown by the black lines with gray bands in
Fig.~\ref{fig:effective-mass-PS}.
We also study the systematic errors by 
investigating the fit range dependence.

%%%%%%%%%%%%%%%%%%%%%%%%%%%%%%%
 %%%%%%%%%%%%%%%%%%%%%%%%%%%%%%%%%%%%%

\begin{figure}[h]
 \begin{center}
        \includegraphics[keepaspectratio, scale=0.7]{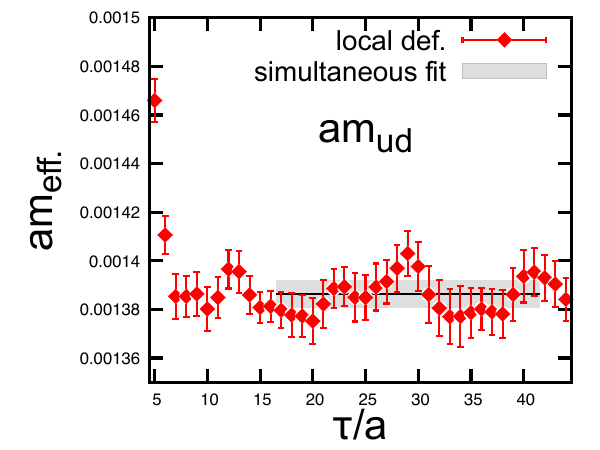}
        \includegraphics[keepaspectratio, scale=0.7]{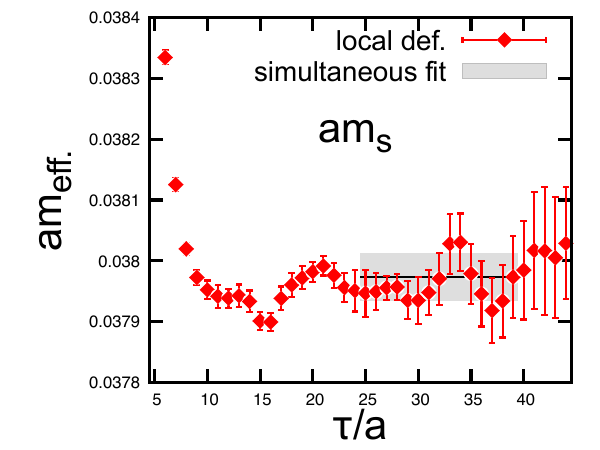}
    \caption{Effective masses for the PCAC bare quark masses for $u,d$ (left) and $s$ (right) quarks
    obtained from the combinations of the PS-PS and PS-A$_4$ correlators 
    in light-light and light-heavy quark sectors.
    The black lines with gray bands depict the results with statistical errors 
    obtained by the simultaneous fit 
    of the PS-PS and PS-A$_4$ correlators.
    }
\label{fig:effective-mass-mq}
\end{center}
  \end{figure}

In Fig.~\ref{fig:effective-mass-mq},
we show the effective PCAC bare quark masses for $u,d$ and $s$ quarks.
The $u,d$ quark mass is obtained from correlators in the light-light quark sector,
while the $s$ quark mass is obtained from the 
combination of correlators in the light-light and light-heavy quark sectors.
Clear plateaux are observed in both quark masses. 
Shown together by the black lines with gray bands 
in Fig.~\ref{fig:effective-mass-mq} are 
the results with statistical errors from the simultaneous fit 
of the PS-PS and PS-A$_4$ correlators combined with Eq.~\eqref{eq:def-mq}.

%%%%%%%%%%%%%%%%%%%%%%%%%%%%%%%
 %%%%%%%%%%%%%%%%%%%%%%%%%%%%%%%%%%%%%

The summary of the results for the PS meson masses and PCAC bare quark masses 
are tabulated in Table~\ref{table:PCAC-PS-data}.
Our result of $m_s/m_{ud} = 27.39(11)(^{+1}_{-0})$ is found to be consistent with
the FLAG average value for $(2+1)$-flavor QCD, 
$m_s/m_{ud} = 27.42(12)$~\cite{FlavourLatticeAveragingGroupFLAG:2021npn}.

%%%%%%%%%%%%%%%%%%%%%%%%%%%%%%%%%%%%%%
%%%%%%%%%%%%%%%%%%%%%%%%%%%%%%%%%%%%%%
\begin{widetext}
\begin{table*}[t]
\centering
\begin{tabular}{ccccc}
  \hline \hline 
$am_{ud}$ & $am_s$ & $m_s/m_{ud}$ & $am_{\pi}$ &  $am_{K}$    \\
$0.001386(6)(^{+1}_{-0}) \quad$ & $ \quad0.037974(40)(^{+2}_{-2}) \quad$ & $ \quad 27.39(11)(^{+1}_{-0}) \quad $& $ \quad0.058606(152)(^{+10}_{-13}) \quad$ & $ \quad0.214552(43)(^{+1}_{-4})$ \\
 \hline \hline 
\end{tabular}
\caption{The results for PCAC (AWI) bare quark masses and PS meson masses
in lattice units.
The errors in the first parentheses are statistical ones
and those in the second parentheses are systematic ones
originating from the fit-range uncertainty of the simultaneous fit.
} \label{table:PCAC-PS-data}
\end{table*}
\end{widetext}
%%%%%%%%%%%%%%%%%%%%%%%%%%%%%%%%%%%%%%
%%%%%%%%%%%%%%%%%%%%%%%%%%%%%%%%%%%%%%

%%%%%%%%%%%%%%%%%%%%%%%%%%%%%%%
 \begin{figure}[htbp]
 \begin{center}
    \includegraphics[keepaspectratio, scale=0.7]{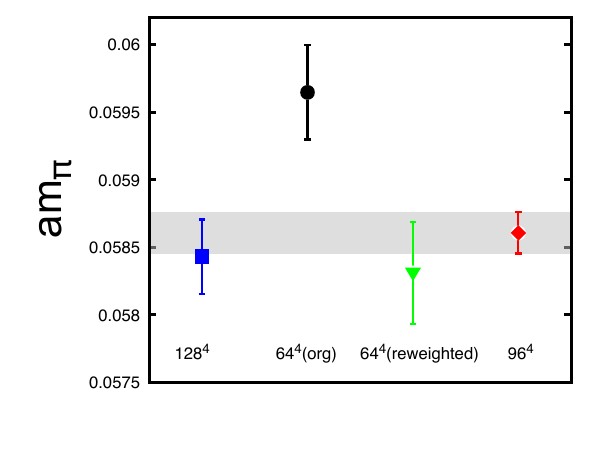}
    \includegraphics[keepaspectratio, scale=0.7]{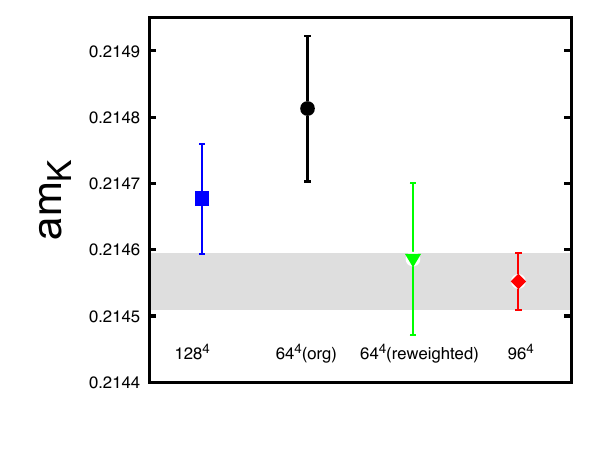}
    \caption{Comparison of our results and PACS results~\cite{Ishikawa:2018jee} for pion mass (left) and kaon mass (right) in lattice units, where the error bars are statistical.
    Our results obtained on a $96^4$ lattice are shown in red diamonds
    with gray bands depicting our statistical errors.
    The PACS results on $128^4$, $64^4$ (org) and $64^4$ (reweighted) lattices are 
    shown by blue squares, black circles and green triangles, respectively,
    where $64^4$ (org) and $64^4$ (reweighted) denote results at the simulation point
    and those at the reweighted point for $(\kappa_{ud}, \kappa_s)$ parameters.
    }
    \label{fig:comp-mass-PS}
\end{center}
\end{figure}
 %%%%%%%%%%%%%%%%%%%%%%%%%%%%%%%%%%%%%

%%%%%%%%%%%%%%%%%%%%%%%%%%%%%%%
 \begin{figure}[htbp]
 \begin{center}
    \includegraphics[keepaspectratio, scale=0.7]{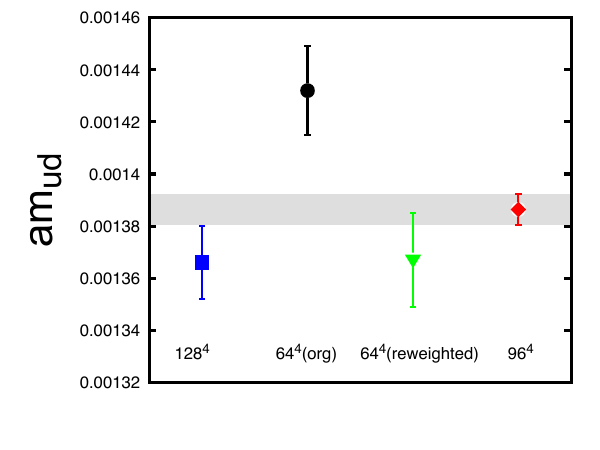}
    \includegraphics[keepaspectratio, scale=0.7]{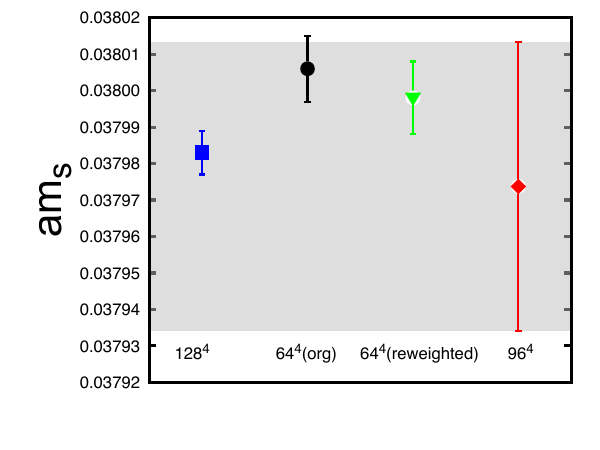}
    \caption{
    Same as Fig.~\ref{fig:comp-mass-PS}, 
    but for PCAC (AWI) bare quark masses for $u,d$ (left) and $s$ (right) quarks
    in lattice units.
    }
    \label{fig:comp-mass-mq}
\end{center}
\end{figure}
 %%%%%%%%%%%%%%%%%%%%%%%%%%%%%%%%%%%%%

We also compare our numerical results on a $96^4$ lattice
with those given by the PACS Collaboration
on $128^4$ and $64^4$ lattices~\cite{Ishikawa:2018jee, PACS:2019ofv}, 
since we take the same simulation parameters, ($\beta, \kappa_{ud}, \kappa_s$), with theirs.
We note that, in the case of the PACS results on a $64^4$ lattice,
the results at the simulation point (referred to $64^4$ (org))
correspond to those at slightly different physical quark masses 
due to the finite volume effect on $\kappa_c$,
so the results at the reweighted $(\kappa_{ud}, \kappa_s)$ parameters
(referred to $64^4$ (reweighted)) are the relevant ones to be compared with~\cite{Ishikawa:2018jee}.

Figure~\ref{fig:comp-mass-PS} shows the comparison for the pion and kaon masses in lattice units.
Our results are consistent with the PACS results on $128^4$ and $64^4$ (reweighted) lattices within  $\sim 1 \sigma$ error, and have smaller statistical errors thanks to higher statistics.
We note that the estimate based on the one-loop SU(3) chiral perturbation theory (ChPT)
shows that the finite volume effects are much smaller than
the statistical errors in these results~\cite{Colangelo:2005gd, Ishikawa:2018jee}.

In Fig.~\ref{fig:comp-mass-mq}, we show the comparison
for the PCAC bare quark masses %, $m^{\rm AWI}$, 
of $u, d$ and $s$ quarks in lattice units.
The obtained $u, d$ quark mass is consistent with the PACS result on $128^4$ and $64^4$ (reweighted) lattices, and our result has a smaller error as expected.
On the other hand, 
the result of the $s$ quark mass exhibits unexpected behavior.
While the results are consistent with each other within the errors, 
our result has a much larger statistical error than those of the PACS result.

It turns out that different flavor sector of the correlators are used to extract $s$ quark mass.
In our case, we use the combination of light-light and light-heavy quark sector 
of correlators, Eqs.~\eqref{eq:PS-PS} and ~\eqref{eq:PS-A4},
while the PACS Collaboration used those only in the {\it heavy-heavy} quark sector
(without so-called disconnected diagrams)~\footnote{N. Ukita, private communication}.
In fact, we checked that the statistical error of the $s$ quark mass from our data becomes 
smaller than those of the PACS results,
{\it if} we use the heavy-heavy quark sector of correlators.

We, however, note that the 
PCAC relation holds for the flavor non-singlet sector,
and using heavy-heavy quark sector in the $(2+1)$-flavor QCD introduces 
systematic errors 
associated with  
the disconnected diagrams
and/or the (partially) quenching effects 
for the heavy valence quarks.
The renormalization as well as the ${\cal O}(ma)$ improvement becomes also non-trivial 
in this case.
Given that the %(relative magnitude of) 
statistical error of $s$ quark mass 
obtained from the heavy-heavy sector
becomes extremely small, ${\cal O}(10^{-4})$~\cite{Ishikawa:2018jee},
the systematic errors mentioned above may not be neglected.
Therefore, we use the combination of
flavor off-diagonal light-light and light-heavy sectors for the $s$ quark mass.

In the analysis by the PACS Collaboration, 
a large finite volume effect on $\kappa_c$ on a $64^4$ lattice was indicated
from the comparison of the PS meson masses and the quark masses between 
$128^4$ and $64^4$ (org)~\cite{Ishikawa:2018jee}.
Since our results on $96^4$ are found to be consistent
with PACS results on $128^4$, we conclude that
$\kappa_c$ is also consistent between $96^4$ and $128^4$.

\subsection{Numerical results: decay constants of pseudo-scalar mesons}
The last quantities obtained from the simultaneous fit of the PS-PS and PS-A$_4$ correlation functions are the decay constants for pion and kaon.
The obtained values are summarized in Table~\ref{table:fpi-fK}.
%%%%%%%%%%%%%%%%%%%%%%%%%%%%%%%%%%%%%%
%%%%%%%%%%%%%%%%%%%%%%%%%%%%%%%%%%%%%%
\begin{widetext}
\begin{table*}[h]
\centering
\begin{tabular}{ccc}
  \hline \hline 
  $af_\pi$  & $af_K$ & $f_K/f_\pi$   \\
$ 0.056980(136)(^{+4}_{-6}) ( 690)  \quad$  & $ \quad  0.067792(68)(^{+5}_{-5}) (822) \quad$  & $  \quad 1.1898(26)(^{+0}_{-1}) \quad$  \\
 \hline \hline 
\end{tabular}
\caption{
The results for the decay constants for pion and kaon in lattice units.  
The errors in the first and second parentheses are statistical ones
and systematic ones originating from the fit-range uncertainty of the simultaneous fit, respectively.
The systematic errors given in the third parentheses for $f_\pi, f_K$ are estimated
from uncertainties of $Z_A$. 
} \label{table:fpi-fK}
\end{table*}
\end{widetext}
%%%%%%%%%%%%%%%%%%%%%%%%%%%%%%%%%%%%%%
%%%%%%%%%%%%%%%%%%%%%%%%%%%%%%%%%%%%%%

\begin{figure}[h]
 \begin{center}
    \includegraphics[keepaspectratio, scale=0.7]{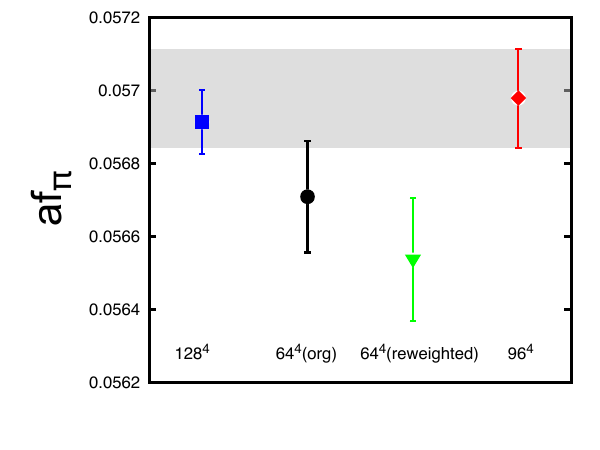}
    \includegraphics[keepaspectratio, scale=0.7]{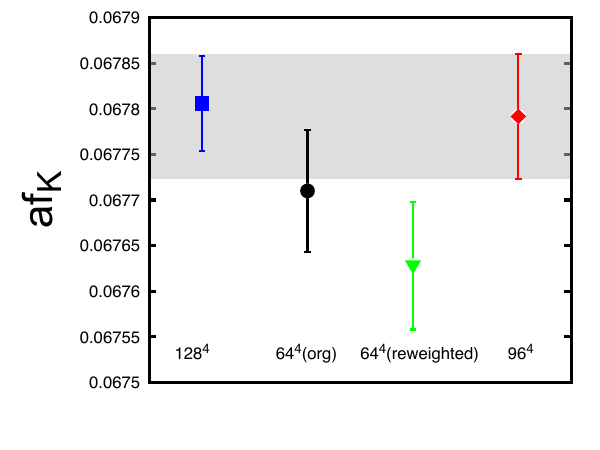}
    \caption{
    Same as Fig.~\ref{fig:comp-mass-PS}, 
    but for decay constants of pion (left) and kaon (right) in lattice units.}
    \label{fig:comp-fpi-fK}
\end{center}
\end{figure}

The comparison plots between ours and the PACS results
with statistical errors
are shown in Fig.~\ref{fig:comp-fpi-fK}.
Note that systematic errors originating from 
uncertainties of the renormalization constant, $Z_A$,
are irrelevant in this comparison,
since $Z_A$ should be common in both results.
We observe that our data are consistent with those from the PACS Collaboration,
in particular with the PACS results on a $128^4$ lattice.
In fact, SU(3) ChPT indicates that the finite volume effects are much smaller
than the statistical errors 
in our results on $96^4$ and PACS results 
on $128^4$~\cite{Colangelo:2005gd, Ishikawa:2018jee}.
On the other hand,
the PACS results on $64^4$ (reweighted)
show some
discrepancies from our results on $96^4$ and PACS  results on $128^4$.
The reason is most likely the finite volume effect
on a $64^4$ (or $(5.4 {\rm fm})^4$) lattice,
as is semiquantitatively supported
by the estimates from SU(3) ChPT~\cite{Ishikawa:2018jee}.

%%%%%%%%%%%%%%%%%%%%%%%%%%%%%%%
 %%%%%%%%%%%%%%%%%%%%%%%%%%%%%%%%%%%%%

Finally, we compare our results of 
$f_K/f_\pi = 1.1898(26)(^{+0}_{-1})$
with the FLAG average value of $f_{K^\pm}/f_{\pi^{\pm}}$.
To this end, we follow the prescription to estimate
the strong-isospin correction based on 
NLO SU(3) ChPT~\cite{FlavourLatticeAveragingGroupFLAG:2021npn},
\beq
\frac{f_{K^\pm}}{f_{\pi^\pm}} &=& \frac{f_K}{f_\pi} \sqrt{1+ \delta_{\rm SU(2)}}, \\
         \delta_{\rm SU(2)}& \approx&
        \sqrt{3}\,\epsilon_{\rm SU(2)}
        \left[-\frac{4}{3} \left(f_K/f_\pi-1\right)+\frac 2{3 (4\pi)^2 f_0^2}
        \left(m_K^2-m_\pi^2-m_\pi^2\ln\frac{m_K^2}{m_\pi^2}\right)
        \right]\,.
\eeq
Using $\epsilon_{\rm SU(2)} = \sqrt{3}/(4R)$ with $R=38.1(1.5)$,
$F_0=f_0/\sqrt{2} = 80(20)$ MeV~\cite{FlavourLatticeAveragingGroupFLAG:2021npn} 
and the values of $m_\pi, m_K$ determined in Sec.~\ref{sec:scale-setting},
we estimate
%\beq
%$\delta_{\rm SU(2)} = -0.0038(^{+11}_{-06})$
$\delta_{\rm SU(2)} \simeq -0.0038$
%\eeq
and obtain  
\beq
\frac{f_{K^\pm}}{f_{\pi^\pm}} &=& 1.1875(26)(^{+0}_{-1}) (23),  
\eeq
where the error in the third parenthesis is due to 
100\% uncertainty assumed for $\delta_{\rm SU(2)}$.
This result is found to be consistent with the FLAG average value for 
$(2+1)$-flavor QCD, 
$f_{K^\pm}/f_{\pi^{\pm}} = 1.1917(37)$~\cite{FlavourLatticeAveragingGroupFLAG:2021npn}
within uncertainties.

%%%%%%%%%%%%%%%%%%%%%%%%%%%%%%%%%%%%%%
%%%%%%%%%%%%%%%%%%%%%%%%%%%%%%%%%%%%%%
\section{Stable Baryons}
\label{sec:baryon-corr}

%%%%%%%%%%%%%%%%%%%%%%%%%%%%%%%%%%%%%%
%%%%%%%%%%%%%%%%%%%%%%%%%%%%%%%%%%%%%%
In this section, 
we calculate the masses
of the five baryons,
$N$, $\Lambda$, $\Sigma$, $\Xi$, and $\Omega$,
which are low-lying stable baryons 
in strong interaction. 
The obtained result for the $\Omega$ mass will be used for the scale setting
in Sec.~\ref{sec:scale-setting}.

%------------------------------------%
\subsection{Calculation strategy}
\label{subsec:baryon-strategy}
%------------------------------------%

The mass of a baryon $B$ is calculated
from two-point temporal correlators,
\beq
C(\tau)
=
\sum_{{\vec x}}\langle 
%\mathcal{J}
B^{\rm (sink)}({\vec x},\tau) 
%\overline{\mathcal{J}}
\overline{B}^{\rm (src)}(\vec{y}, 0) 
\rangle,  \label{eq:CorrBaryon}
\eeq 
where we omit the spin indices for simplicity.

The baryon operator $B(x)$ with $x=(\vec{x},\tau)$ for an octet baryon %with spin 1/2 
is defined by 
\beq
B(x)=\epsilon^{abc}
(q_{a}(x)^TC\gamma_5 q_{b}(x))q_{c}(x), 
\eeq
where $C=\gamma_4\gamma_2$ is the charge conjugation matrix
and $a,b,c$ are the color indices. 
The spin and parity projection to $J^P=\frac{1}{2}^+$
is performed as
\beq
C^{({\rm O})}(\tau)
= {\rm Tr}\left[ P^{\frac{1}{2}^+} C(\tau)
\right]
\eeq
with
\beq
P^{\frac{1}{2}^+} \equiv \frac{1+\gamma_4}{2} ,
\eeq
where ${\rm Tr}$ is taken in the spinor space.

For the decuplet baryons with spin $3/2$,
we employ
the correlator of the Rarita-Schwinger fields defined as 
\beq
C_{ij}(\tau)
&=&
\sum_{{\vec x}}\langle 
%\mathcal{J}
B^{\rm (sink)}_i({\vec x},\tau) 
%\overline{\mathcal{J}}
\overline{B}^{\rm (src)}_j(\vec{y}, 0) 
\rangle, \\
B_i(x) &=& \epsilon^{abc}
(q_{a}(x)^TC\gamma_i q_{b}(x))q_{c}(x), 
\eeq
where we explicitly show the 
indices for 
the spatial label of the 
Rarita-Schwinger fields,  
$i, j =1,2,3$.
The spin and parity projection to $J^P=\frac{3}{2}^+$
is performed as
\beq
C^{({\rm D})}(\tau)
=
\sum_{
\substack{i,j=1,2,3}}
{\rm Tr}\left[ 
P_{ij}^{\frac{3}{2}^+}
C_{ij}(\tau)
\right]
\eeq
with
\beq
P_{ij}^{\frac{3}{2}^+}
\equiv
\frac{1+\gamma_4}{2}(\delta_{ij}-\frac{1}{3}\gamma_i\gamma_j).
\eeq

In the calculation of the correlators,
there is additional freedom to introduce 
smearing in the sink and/or source quark operators.
After benchmark of several possibilities,
we employ the analysis
of the correlators with the point-sink and wall-source operators
as our main results, 
where
the quark operator in the source operator
$B^{\rm (src)}$ is replaced with 
the wall-source one,
\beq
q^{({\rm w})}_{a}(\tau)\equiv
\sum_{\vec y}q_{a}({\vec y}, \tau),
\label{eq:OpWallSourceQuark}
\eeq
together with the Coulomb gauge fixing for the configurations.
A reason for this choice is that
the large amount of statistical data is 
accumulated for this setup as the by-product of our
calculations of two-baryon systems with the wall-source,
and we can achieve better precision
compared to other choices.

In the case of the mass of $\Omega$, which 
is obtained with the highest precision and 
is used 
for the scale setting, % in our study,
we perform additional calculations
using the variational method~\cite{Luscher:1990ck}
with the smeared-sink and smeared-source operators.
In this method, a rigorous upper bound for the mass can be obtained,
and we can estimate the systematic error 
in the result obtained from the wall-source.

%[formalism of variational methods]
In the variational method,
a generalized eigenvalue problem (GEVP) is solved 
for the correlator matrix of the baryon.
The $(l,l')$-element of the correlator matrix is defined as 
the two-point temporal correlators 
with $l$-th smeared-sink and $l'$-th smeared-source operators
\beq
\left[ C(\tau) \right]_{ll'} = 
\sum_{\vec{x}}\langle B^{(l)} (\vec{x}, \tau)
\overline{B}^{(l')}(\vec{y}, 0) \rangle.
\label{eq:CorrSmearedSource}
\eeq
Here, 
the smeared baryon operators, 
$B^{(l)}$,
are given by
replacing the local quark operator 
with the smeared one, %quark operator
\beq
q^{(l)}_{a}({\vec x},\tau)
=
\sum_{\vec y}f_{q}^{(l)}(|{\vec x}-{\vec y}|)q_{a}({\vec y},\tau),
\label{eq:q-smear}
\eeq
where $f_{q}^{(l)}(r)$ is a smearing function.

The GEVP of the correlator matrix is solved as
\beq
\tilde{C}(\tau)=V^\dagger(\tau_D)C^{-1/2}(\tau_0)C(\tau)C^{-1/2}(\tau_0)V(\tau_D) 
\label{eq:GEVP}
\eeq
with the parameter $\tau_0$ and $\tau_D$, 
and the rotational matrix $V(\tau_D)$ 
is defined to make the matrix 
$\tilde{C}(\tau_D)$
diagonal.
It is also checked that $\tilde{C}(\tau)$ at $\tau\neq \tau_D$
is almost diagonal.
The component $[\tilde{C}(\tau)]_{11}$ corresponds to 
the correlator of the lowest energy state. 
The parameters $\tau_0$ and $\tau_D$ with 
$\tau_D > \tau_0 \geq \tau/2$~\cite{Blossier:2009kd}  should be large to suppress 
the contribution from the excited states. 

For the calculation of the correlators
with smeared-sink and smeared-source,
we utilize the $Z_3$ noise 
source method 
\cite{PhysRevD.82.114501, PhysRevD.88.014503, Wu_2018, PACS:2019ofv}
in order to reduce the statistical fluctuations.
The three-dimensional lattice of size $L^3$ is divided into $n_{\rm sub}^3$ sub-lattices 
with the size of $L_{\rm sub}^3=(L/n_{\rm sub})^3$,
and the sub-lattices are labeled by the even-odd coloring.
The spatial center of the smearing function $f_q^{(l)}$
is located on each even-colored sub-lattice
so that $(n_{\rm sub}^3/2)$ copies of the original smearing function 
are distributed in the total lattice,
and each copy of the smearing function is multiplied
by the $Z_3$ random number.
In this way, we can prepare a quark source operator
which has support for an independent single-baryon operator 
at $(n_{\rm sub}^3/2)$ spatially different places simultaneously.
The spatial distribution of this work
is same as the $s2$ (even-odd) spatial dilution 
(in sub-lattice degrees of freedom) given in Ref.~\cite{Akahoshi:2019klc},
and also as 
the crystal structure of face-centered-cubic (FCC) in Ref.~\cite{PACS:2019ofv}.

In our study,
we employ
a Gaussian-type smearing function.
Moreover, we use the tail-cut technique for the smearing,
where the smearing function is defined within a finite distance
in order to avoid overlap of the 
smearing functions from spatially separated sources. 
The explicit form for the smearing function 
in Eq.~(\ref{eq:q-smear})
is given as
\beq
%\begin{align}
f_{q}^{(l)}(r)=
\begin{cases}
A_{q}^{(l)}{\rm e}^{-\alpha_{q}^{(l)} r^2} 
\quad &\text{for} \ 0 < r < r_{\rm cut}, \\
1 \quad &\text{for} \ r=0, \\
0 \quad &\text{for} \ r_{\rm cut} \leq r.
\end{cases}
\label{eq:smear}
\eeq
%\end{align}
%
When combined with the $Z_3$ noise source method
on our lattice with $L/a=96$,
we take $L_{\rm sub}/a = 24$ with $n_{\rm sub}=4$
so that $n_{\rm sub}^3/2 = 32$ single baryon sources
are distributed on the lattice
(i.e.,
at spacial locations of
$(0, 0, 0),
(48, 0, 0),
(24, 24, 0),
(72, 24, 0),
(0, 48, 0),
(48, 48, 0),
(24, 72, 0),
(72, 72, 0),
(24, 0, 24),
\cdots,
(48,72,72)
$),
and the tail-cut parameter is taken as 
$r_{\rm cut} = L_{\rm sub}/(\sqrt{2}a)$. %L_sub*sqrt{2}/2
We confirm that 
both the $Z_3$ noise method and 
the tail-cut technique 
improve the signal-to-noise 
of the results of the 
baryon masses.

We have performed benchmark calculations with relatively low statistics
to tune the dimension of the correlator matrix 
(from $2\times 2$ to $5 \times 5$)
and the smearing parameters.
In general, 
the lowest eigenstate of the correlator matrix
is expected to have a larger overlap with the desired ground state 
if the matrix has a larger size. 
However, we found that the statistical error increases 
with the size of the correlator matrix. 
Finally, we employ the $2\times2$ correlator matrix 
in our study with high statistics
to obtain the clearest signal of the $\Omega$ baryon mass.
The corresponding two sets of the smearing parameters 
for the $s$-quark sector $(q=s)$
are taken as 
$A_{s}^{(1)} = A_{s}^{(2)} = 1$,
$(\alpha_s^{(1)}, \alpha_s^{(2)})
= (0.30250, 0.02560)$,
where 
the smearing parameters of $l=1,2$ 
correspond to the narrow and broad extent 
of the quark fields, respectively.
The mass of $\Omega$ is obtained by solving GEVP
of the corresponding $2\times 2$ correlator matrix.

We also performed supplemental calculations for other 
baryons in the $2\times 2$ variational method
using the smearing parameters for the $u, d$-quark sectors,
$A_{u,d}^{(1)} = A_{u,d}^{(2)} = 1$,
$(\alpha_{u,d}^{(1)}, \alpha_{u,d}^{(2)})
= (0.13505, 0.01155)$,
whereas the parameters for the $s$-quark are taken to be
the same as in the $\Omega$ case.
However, we found that the statistical fluctuations are much
larger than the case of $\Omega$. Considering 
that the computational cost is also much larger due to 
the $u,d$-quark solver, we do not use the variational method
for baryons other than $\Omega$.

%------------------------------------%
\subsection{Numerical results: baryon masses}
\label{subsec:baryon-results}
%------------------------------------%
%%% introduce effective mass
First, we present analyses of the masses of baryons 
from the correlators with the point-sink and wall-sources.
Quark propagators are solved 
with the periodic boundary condition for all directions
%where
and 
stopping conditions 
$|\rcrit| < 10^{-8} (10^{-12})$ for $u,d$-quarks ($s$-quark).
%%% About statistics
The hypercubic symmetry on the lattice 
(4 rotations and 96 temporal source locations) 
is used
with 1600 configurations
as well as the average of the forward/backward propagations,
so that
the total number of measurements is 
$1600\times 4 \times 96 \times 2 = 1,228,800$. 
The statistical errors are estimated by the jackknife method 
with 20 jackknife samples (${\rm bin~size}=80$ configurations). 
We have checked that the results are almost unchanged 
in the case of ${\rm bin~size}=160,40,20$ configurations.

%--- 5-baryon effective mass plots, summary ---%
%%%%%%%%%%%%%%%%%%%%%%%%%%%%%%%
 \begin{figure}[htbp]
 \begin{center}
        \includegraphics[keepaspectratio, scale=0.5]{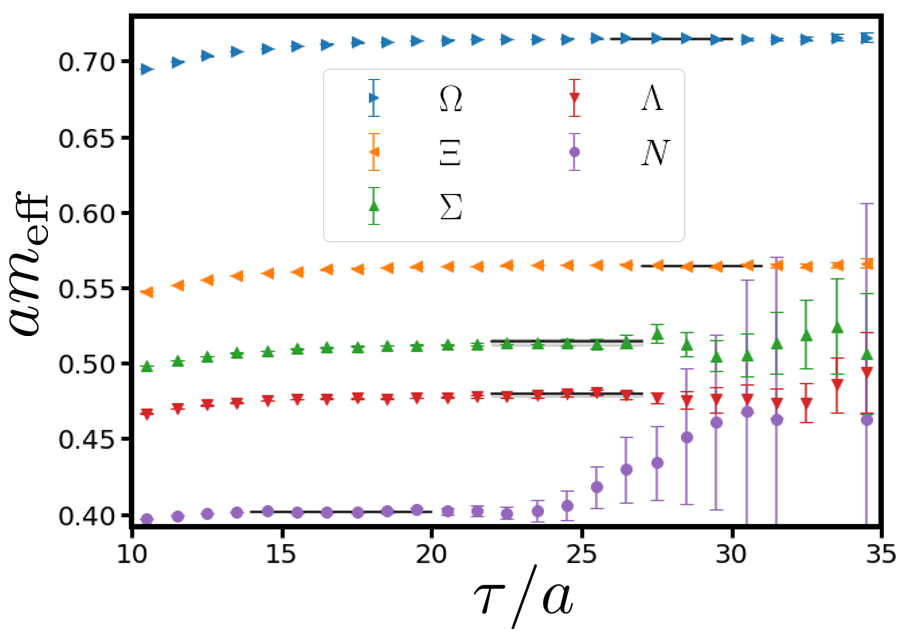}
    \caption{The effective masses for five baryons in lattice units.
    The fit range and fit results with statistical errors
    are shown by black lines with gray bands.
}\label{fig:effective-mass-5baryons}
\end{center}
  \end{figure}
 %%%%%%%%%%%%%%%%%%%%%%%%%%%%%%%%%%%%%
 
%--- 5 each baryon effective mass plots ---%
%%%%%%%%%%%%%%%%%%%%%%%%%%%%%%%
 \begin{figure}[htbp]
 \begin{center}
        \includegraphics[keepaspectratio, scale=0.40]{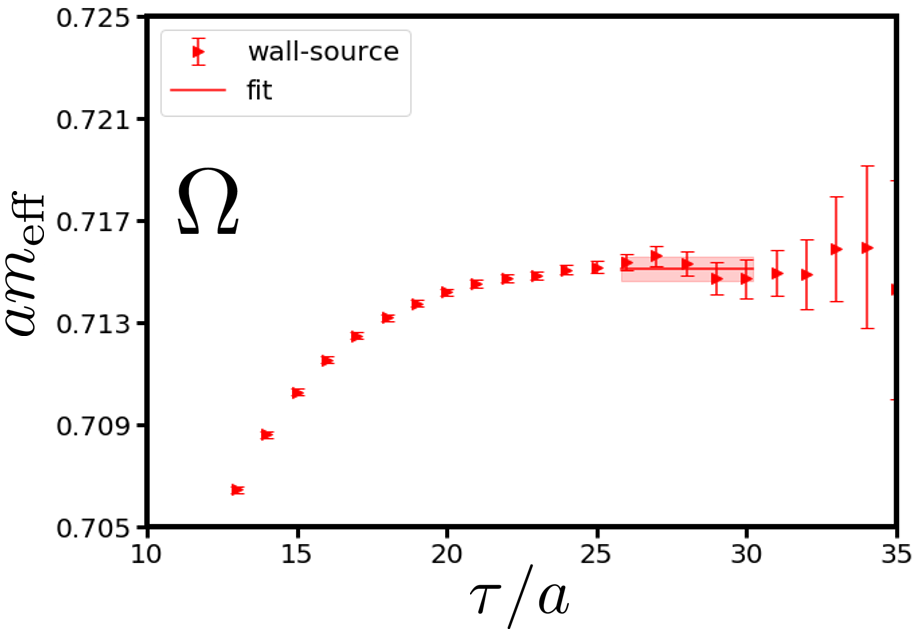}       \includegraphics[keepaspectratio, scale=0.40]{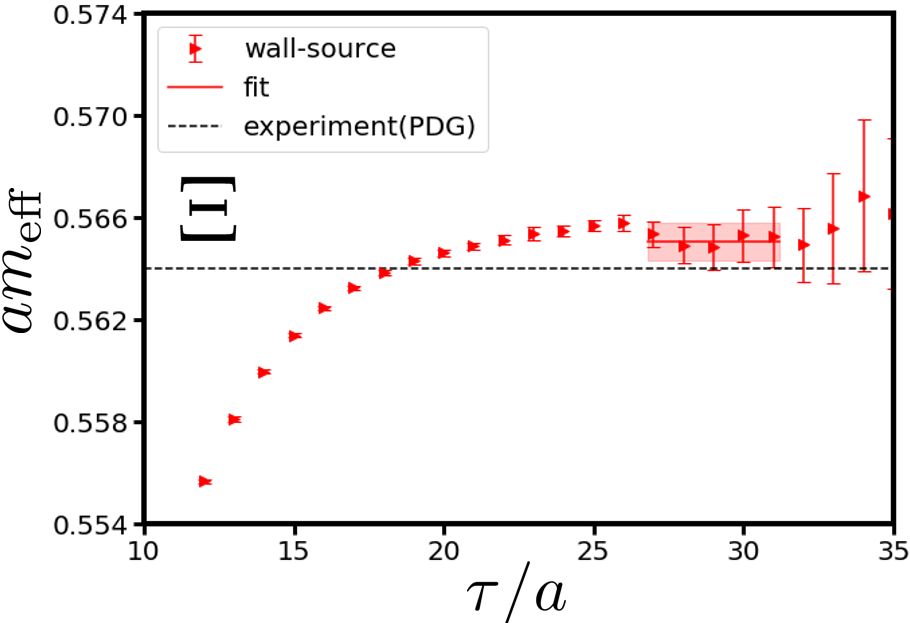}\\
        \includegraphics[keepaspectratio, scale=0.40]{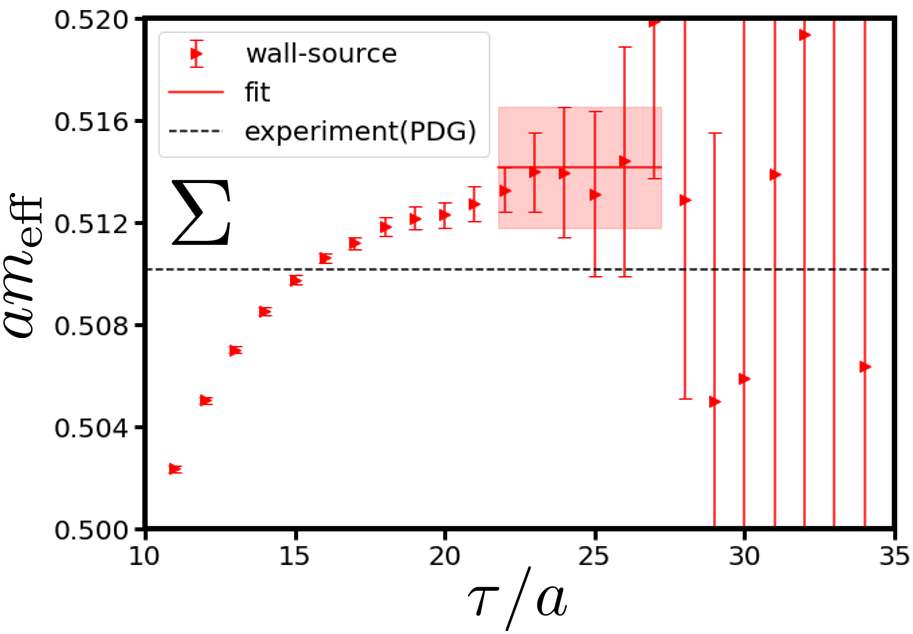}
        \includegraphics[keepaspectratio, scale=0.40]{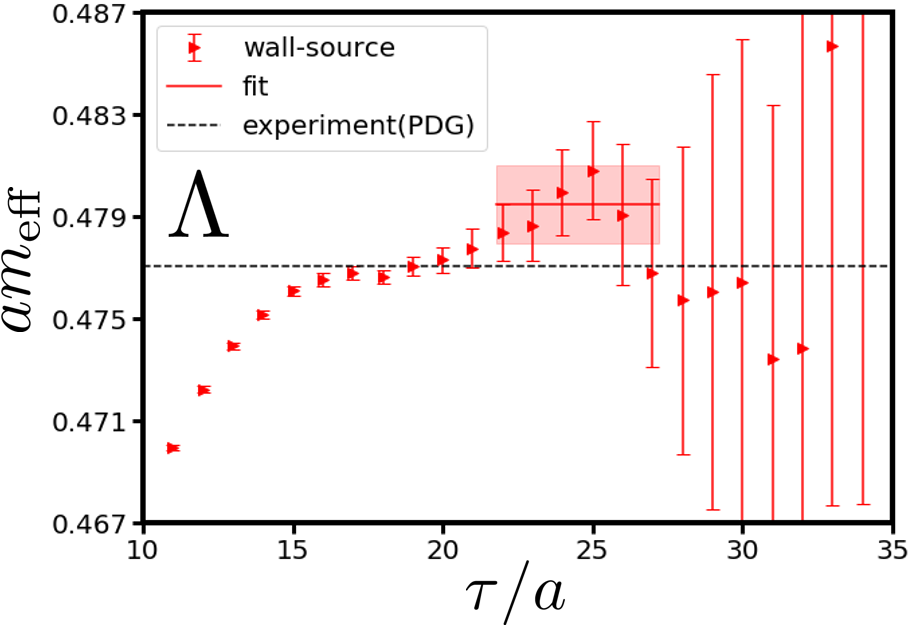}\\        \includegraphics[keepaspectratio, scale=0.40]{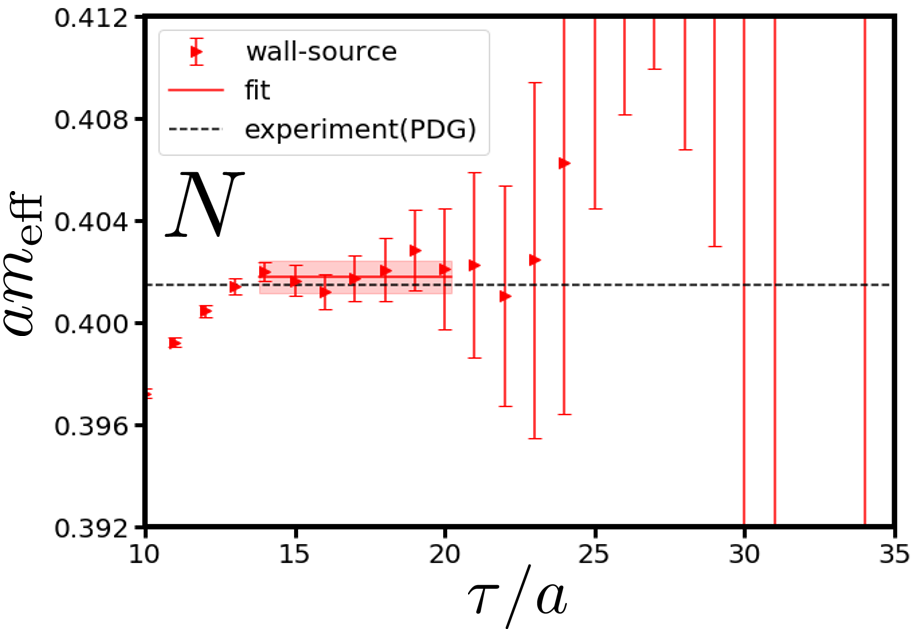}
    \caption{
    Effective masses for five low-lying baryons with point-sink and wall-source correlators in lattice units.
    The red lines are the fit results with the red bands denoting the statistical errors.
    The dotted lines denote the isospin-averaged experimental results from Particle Data Group(PDG)~\cite{Workman:2022ynf}, where the $\Omega$ mass is used for the scale setting.
}\label{fig:each-baryon}
\end{center}
  \end{figure}
 %%%%%%%%%%%%%%%%%%%%%%%%%%%%%%%%%%%%%

In Fig.~\ref{fig:effective-mass-5baryons},
we show the effective masses of five baryons simultaneously,
and the enlarged figures for each baryon are shown in Fig.~\ref{fig:each-baryon},
where
the effective mass is defined as
\beq
m_{\rm eff}(\tau)\equiv \log\frac{C^{(B)}(\tau)}{C^{(B)}(\tau+1)}
\label{eq:meff:wall}
\eeq
with $C^{(B)}(\tau) = C^{(O)}(\tau)$ or $C^{(D)}(\tau)$.
We observe a good plateau for each baryon corresponding to the ground state saturation.

The baryon masses are obtained
by fitting the correlators by a single exponential functional form
with the fit ranges chosen from 
the plateau regions in the effective mass plots. 
In Figs.~\ref{fig:effective-mass-5baryons} and \ref{fig:each-baryon},
the fit results are shown by
black lines with gray bands denoting the statistical errors. 
The numerical fit results with statistical errors 
and the fit ranges are summarized in Table \ref{table:FitBaryonsMassLattice}. 
We also consider several different choices for the fit ranges
by changing the lower and/or upper bounds of ranges
by several time slices,
and estimate the systematic errors in the baryon masses.
The resulting systematic errors for the octet baryons 
are given in Table \ref{table:FitBaryonsMassLattice}. 
The relative magnitudes of the finite volume effect are expected to be
$\sim (m_{\rm PS}/m_B)\cdot e^{-(m_{\rm PS}L)}/(m_{\rm PS}L) < {\cal O}(10^{-4})$
from ChPT with $m_{\rm PS} (m_B)$ being a relevant PS meson mass (baryon mass)~\cite{Gasser:1987zq}, 
and thus are negligible in the current precision.
In the case of the $\Omega$ mass, the systematic errors
are estimated not only from the fit range dependence
but also from the analysis with the variational method
as will be described below.
This additional study is performed because the $\Omega$ mass is determined 
with the highest precision and is used for the scale setting.

%--- 5-baryon's mass table, summary ---%
\begin{table}[hbtp]
  \caption{
  The results of %the fitting analysis for 
  the masses of the five baryons in lattice unit. 
  The central values as well as statistical errors
  (in the first parentheses)
  are obtained from the analyses of the wall-source data
  with the fit ranges given in the table.
  The systematic errors (in the second parentheses)
  are estimated from the 
  fit range dependence.
  In the case of the $\Omega$ mass,
  the result of the variational method 
  is additionally used to estimate the systematic error.}
  \label{table:FitBaryonsMassLattice}
  \centering
  \begin{tabular}{clc}
    \hline
    baryon & ~~~~~mass & fit range \\
    \hline %\hline
    $N$        & $0.40179(64)(^{+4}_{-20})$ & [14,20]\\
    $\Lambda$  & $0.47947(154)(^{+18}_{-95})$ & [22,27]\\
    $\Sigma$   & $0.51414(237)(^{+11}_{-105})$ & [22,27]\\
    $\Xi$      & $0.56469(74 )(^{+58}_{-0})$ & [27,31]\\
    \hline
    $\Omega$   & $0.71510(46 )(^{+93}_{-5})$& [26,30]\\
    \hline
  \end{tabular}
\end{table}

%%% Results of variational method
We present the details of the analysis with the variational method
for the $\Omega$ baryon.
The $2\times 2$ correlator matrix is calculated
with the $Z_3$ noise method and the tail-cut technique
at 16 temporal source locations,
and thus the total number of measurements is
$1600\ \mbox{confs} \times 4\ \mbox{rotations}\times 16\ \mbox{sources} \times 2\ 
\mbox{propagations} = 204,800$. 
The statistical errors are estimated by the jackknife method 
with ${\rm bin~size}=80$ configurations as in the case of 
the wall-source analysis
and bin size dependence is found to be negligible.
In terms of the parameters in the variational method, 
$\tau_0$ and $\tau_D$ in Eq.~(\ref{eq:GEVP}),
we tried several combinations
and  
$(\tau_0,\tau_D)=(19,22)$ is found to give 
the clearest signal %of the effective mass plot 
for the $\Omega$ baryon. 

\begin{figure}[htbp]
 \begin{center}
        \includegraphics[keepaspectratio, scale=0.50]{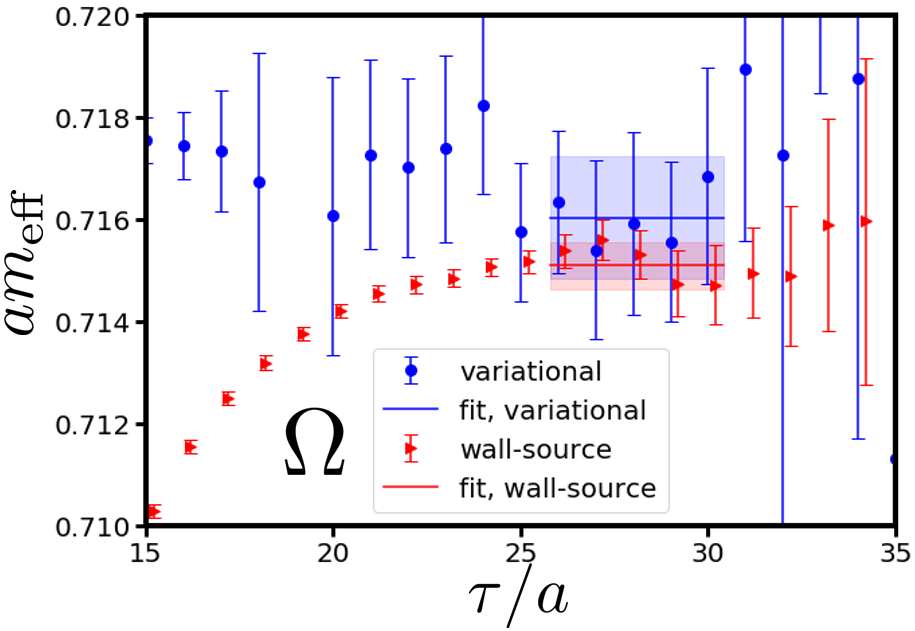}
    \caption{The effective mass for $\Omega$ baryon in lattice units with two analyses, one from the point-sink and wall-source operators (red triangles) 
    and the other from the variational method (blue circles).
    The fit results as well as the corresponding statistical errors 
    for each analysis
    are shown by lines and bands, respectively.
    The data are slightly shifted horizontally for visibility. 
}\label{fig:effective-mass-Omega}
\end{center}
  \end{figure}

The effective mass of $\Omega$
with the variational method is shown 
in Fig. \ref{fig:effective-mass-Omega}
by blue circles.
Here, the effective mass is defined by
\beq
\tilde{m}_{\rm eff}(\tau)
=-\frac{1}{\tau-\tau_0}\log[\tilde{C}(\tau)]_{11}
\label{eq:meff:variational}
\eeq
with $[\tilde{C}(\tau)]_{11}$ being the lowest diagonal element of 
Eq.~(\ref{eq:GEVP}).
This definition is found to give smaller statistical errors
than the case of Eq.~(\ref{eq:meff:wall}).
Shown together in Fig.~\ref{fig:effective-mass-Omega}
by red triangles 
are the effective masses from the point-sink and wall-source
(with the definition of Eq.~(\ref{eq:meff:wall}) as before).
We find that 
the results 
from the variational method and the wall-source
converge to consistent plateaux within the error bars 
around $26\lesssim \tau/a \lesssim 30$.
By fitting $[\tilde{C}(\tau)]_{11}$ with a single exponential function,
we obtain the result of the variational method as 
$am_{\Omega}^{({\rm var.})}=0.71603(120)$
with the fit range $[26,30]$.
The result is insensitive to a change of the fit range.

%%% reason of diff of wall and var. method.
We note that 
the result from the variational method 
gives us the upper bound of the true value, 
while 
that
from the wall-source analysis does not have 
such a property. 
In our analysis, 
we see that %the upper bound of 
the $\Omega$ mass 
from the variational method decreases
as the parameters $(\tau_0,\tau_D)$ increase.
However, the statistical error is found to become larger in the case of 
larger $(\tau_0,\tau_D)$. 
This is why we employ 
$(\tau_0,\tau_D) = (19,22)$
as an optimal choice
in the variational method
and we utilize the corresponding result 
to estimate the upper side of the systematic error for the $\Omega$ mass.

Our final result for the $\Omega$ mass leads to
$am_{\Omega} = 0.71510(46 )(^{+93}_{-5})$
and is given in Table \ref{table:FitBaryonsMassLattice}. 
The central value and statistical error are obtained 
from the wall-source analysis.
The systematic error is estimated from the largest deviation
considering both
the upper bound given from the variational method
and the fit range dependence in the wall-source analysis.

%%% usefulness of wall-source, p-wave?
In the study of the PACS Collaboration, which employs essentially 
the same configuration setup as this study,
any reasonable plateau is not found for $\Omega$
in their point-sink and (exponential-type) smeared-source two-point correlator~\cite{PACS:2019ofv}.
A possible reason why
we observe a clear plateau in our point-sink and 
wall-source correlator
is that 
excited state contaminations from meson-baryon scattering states 
are expected to be suppressed  in the wall-source
because the dominant meson-baryon states are the P-wave states. 
We also note that our statistics are much larger than
those in Ref.~\cite{PACS:2019ofv}, which is helpful in identifying the plateau.
It could be also interesting to apply 
the generalized pencil of functions analysis for 
the wall-source correlator to further suppress excited state contaminations~\cite{Hudspith:2024kzk}.

In the variational method, such excited state contaminations are 
automatically suppressed by solving the GEVP.
However, we employ only single-baryon (three-quark) type operators in our variational method,
which would have a small overlap with meson-baryon scattering states.
This could be a reason why we have to take large $(\tau_0, \tau_D)$ in our study.
In the future, it will be desirable to employ not only single-baryon type operators
but also meson-baryon type operators in the variational method.

%%%%%%%%%%%%%%%%%%%%%%%%%%%%%%%
 %%%%%%%%%%%%%%%%%%%%%%%%%%%%%%%%%%%%%

%%%%%%%%%%%%%%%%%%%%%%%%%%%%%%%%%%%%%%
%%%%%%%%%%%%%%%%%%%%%%%%%%%%%%%%%%%%%%
\section{Unstable hadrons}\label{sec:unstable}
%%%%%%%%%%%%%%%%%%%%%%%%%%%%%%%%%%%%%%
%%%%%%%%%%%%%%%%%%%%%%%%%%%%%%%%%%%%%%

We present the study of the
low-lying resonances,
the decuplet baryons ($\Delta, \Sigma^*~\rm{and}~\Xi^*$), 
and the vector mesons ($\rho, K^*~\rm{and}~\phi$).
As a rigorous treatment of these hadrons,
it is necessary to calculate the scattering phase shifts
of the relevant decay modes as formulated in
Refs.~\cite{Luscher:1990ux, Ishii:2006ec}
and extract the pole position in the scattering matrix.
Since such a detailed analysis is beyond the scope of this paper,
we here present a simple analysis 
which is performed in the same way as in the case of stable hadrons.
Such an analysis is expected to be a good approximation 
for hadrons with small decay widths, but
large systematic uncertainties are introduced for 
hadrons with large decay widths.

The correlators are calculated 
with the point-sink and wall-source operators
%with the statistics of
%$1600\times 4 \times 96 \times 2 = 1,228,800$ measurements
as described in Sec.~\ref{subsec:baryon-results}.
The effective mass plots 
are shown in Appendix~\ref{sec:unstable-app}.
The fit results are summarized in Table \ref{table:FitResonanceMassLattice}.
We note again that the study considering decay modes is necessary
in the future in particular for hadrons with large decay widths.

%--- resonance hadron's mass table, summary ---%
\begin{table}[hbtp]
  \caption{
  The results of %the fitting analysis for 
  the masses of the resonant hadrons in lattice units. 
  The central values as well as statistical errors
  (in the first parentheses)
  are obtained from the analyses of the wall-source data
  with the fit ranges given in the table.
  The systematic errors (in the second parentheses)
  are estimated from the 
  fit range dependence.
  }
  \label{table:FitResonanceMassLattice}
  \centering
  \begin{tabular}{clc}
    \hline
    hadron &~~~~~mass & fit range \\
    \hline %\hline
    $\Delta$  & $0.54195(105)(^{+7}_{-4})$ & [11,16]\\
    $\Sigma^*$   & $0.60080(417)(^{+115}_{-0})$ & [20,24]\\
    $\Xi^*$      & $0.66178(175)(^{+0}_{-211})$ & [25,28]\\
    \hline
    $\rho$   & $0.31710(8)(^{+36}_{-4147})$& [6,10]\\
    $K^*$    & $0.38156(8)(^{+6}_{-604})$& [11,15]\\
    $\phi$   & $0.43695(8)(^{+1}_{-33})$& [19,25]\\
    \hline
  \end{tabular}
\end{table}

%%%%%%%%%%%%%%%%%%%%%%%%%%%%%%%%%%%%%%
%%%%%%%%%%%%%%%%%%%%%%%%%%%%%%%%%%%%%%
\section{Scale setting and observables in physical units}\label{sec:scale-setting}
%%%%%%%%%%%%%%%%%%%%%%%%%%%%%%%%%%%%%%
%%%%%%%%%%%%%%%%%%%%%%%%%%%%%%%%%%%%%%
% strategy 
The lattice cutoff scale is 
determined in this section. 
In our study, 
we choose the mass of the $\Omega$ baryon, $m_\Omega$,
for the determination of the lattice cutoff scale.
As was shown in Sec.~\ref{subsec:baryon-results},
$m_\Omega$ is determined with the highest precision 
among all the baryons considered
and its  systematic error is thoroughly under control.

Using the experimental value of the $\Omega$ mass~\cite{Workman:2022ynf}, 
$
m_{\Omega}=1672.45 \ [{\rm MeV}], %= m_{\Omega}^{\rm Lat}a,
$
we determine the lattice spacing as 
\beq
a&=0.084372(54)(^{+109}_{-6}) \ [{\rm fm}], \label{eq:a}\\
%a^{-1}
\frac{1}{a}
&=2338.8(1.5)(^{+0.2}_{-3.0}) \ [{\rm MeV}].\label{eq:1overa}
\eeq
We also confirm that the lattice scale 
determined from the $\Xi$ mass is 
consistent with that from the $\Omega$ mass.
Our lattice scale is slightly different from
that obtained from PACS10 configurations~\cite{PACS:2019ofv}, 
which employ essentially the same gauge configuration setup,
$a^{-1}_{\rm PACS10}=2316.2(4.4) \ [{\rm MeV}]$.
For in-depth comparison between two results, see Appendix~\ref{sec:PACS10}.

Using our lattice cutoff scale,
we can now give 
the physical values for 
the renormalized quark masses 
and
the decay constants of PS mesons 
as
\beq
    &&
    m_{ud}^{\overline{\rm MS}} (\mbox{2 GeV}) = 3.225 (^{+16}_{-19}) (46)\ [{\rm MeV}], \quad 
    m_s^{\overline{\rm MS}} (\mbox{2 GeV}) = 88.37 (^{+0.15}_{-0.22}) (1.26)\  [{\rm MeV}], \\
    &&
    f_\pi = 133.3(^{+0.4}_{-0.5}) (1.6)\ [{\rm MeV}], \quad f_K = 158.6 (^{+0.3}_{-0.4}) (1.9)\ [{\rm MeV}] .
\eeq
Here, the first errors denote 
the sum of the statistical and systematic errors 
of each quantity in lattice units
as well as the statistical and systematic errors of the lattice unit
combined in quadrature,
while
the second errors in quark masses and decay constants present
the error of $(Z_A/Z_P)$ and $Z_A$, respectively.

%%%%%%%%%%%%%%%%%%%%%%%%%%%%%%%
 \begin{figure}[htbp]
 \begin{center}
        \includegraphics[keepaspectratio, scale=0.6]{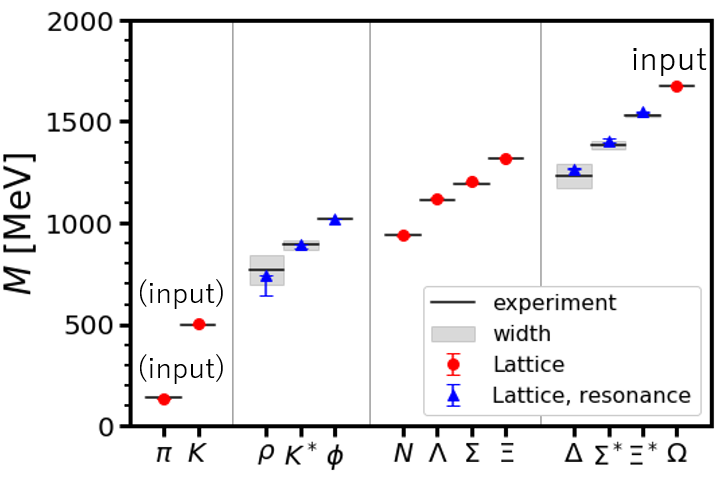}
    \caption{
    Summary of hadron spectrum at the physical point,
    where $m_\Omega$ is used as an input for the scale setting. 
    $m_\pi, m_K$ are outputs in this paper, but 
    they were essentially used as inputs in previous studies
    with the same lattice setting~\cite{Ishikawa:2018jee}.
    Red circles and blue triangles 
    denote the masses of stable hadrons
    and hadronic resonances, respectively,
    with statistical and systematic errors 
    combined in quadrature.
    Horizontal black lines denote the experimental values,
    and gray bands correspond to the decay widths
    of resonances,
    where values are taken from Particle Data Group (PDG)~\cite{Workman:2022ynf} with isospin-averaging.
}\label{fig:summaryfig}
\end{center}
  \end{figure}
 %%%%%%%%%%%%%%%%%%%%%%%%%%%%%%%%%%%%%

The masses of low-lying hadrons in physical units 
are given in Table \ref{table:HadronMassPhysical}. 
The numbers in the first (second) parentheses denote 
the statistical (systematic) errors.
The statistical correlations between 
each hadron and $\Omega$ baryon wall-source correlators
are properly taken into account in the estimate of statistical errors.
In  Table \ref{table:HadronMassPhysical},
the experimental values corresponding to 
the isospin-averaged masses are also shown for comparison.
In Fig.~\ref{fig:summaryfig},
we show a summary plot 
for stable hadrons by red circles
together with experimental values shown by black lines.
In this paper, $m_\Omega$ is used as an input for the scale setting,
and masses of all other hadrons, including $m_\pi, m_K$, are outputs.
In Fig.~\ref{fig:summaryfig}, however,
we indicate that $m_\pi, m_K$ also correspond to inputs,
since quark mass parameters $\kappa_{ud}, \kappa_s$
used in this study were previously tuned
using $m_\pi, m_K$ 
in Ref.~\cite{Ishikawa:2018jee}.

The lattice results of the masses of hadrons are 
found to be sufficiently 
close to the experimental results
considering the uncertainties of 
the isospin breaking effects. 
%in our 2+1 flavor lattice QCD simulations.
This also indicates that the discretization errors
in this study is small,
but 
it would be better to perform 
explicit calculations for the 
continuum extrapolation in the future.
In Fig.~\ref{fig:summaryfig},
the results for hadron resonances are also shown 
by blue triangles.
Note that they are obtained neglecting the effect of hadronic decays
and more proper phase shift studies of relevant decay modes 
are left for future studies.

Finally, we comment on the uncertainties
associated with 
the quark mass tuning and
the ambiguities of the
notion of ``physical point".
While our target values for the physical point are defined by
$m_{\pi^{\rm ave}} = 138.04$ MeV and 
$m_{K^{\rm ave}} = 495.64$ MeV
corresponding to the isospin-averaged experimental values~\cite{Workman:2022ynf},
the outcomes of our simulations for $m_{\pi} (m_K)$ are found to be 
lighter (heavier) by $0.7 (1.2) \%$ than our target values.
In addition,
in the literature of lattice simulations, 
the physical point is 
often defined using 
the experimental values~\cite{Workman:2022ynf} of 
$m_{\pi^0} = 134.98$ MeV and 
$m_{K^0}  = 497.61$ MeV,
as suggested in the FLAG 2021 report~\cite{FlavourLatticeAveragingGroupFLAG:2021npn}.
While a detailed calculation for 
the quark mass tuning with these definitions for the physical point
is beyond the scope of this study,
we estimate how our results would change with 
each definition
(together with 
$m_{\Xi^{\rm ave}}=1318.29$ MeV or
$m_{\Xi^0} = 1314.86$ MeV),
utilizing the quark mass dependence of hadron masses
obtained from the same lattice action as ours
by the PACS Collaboration~\cite{PACS:2019ofv}
(namely, $C_{\rm ud}, C_{\rm s}$ parameters 
in Ref.~\cite{PACS:2019ofv}).
It turns out that,
in the case of 
the definition with 
$(m_{\pi}, m_K, m_{\Xi}) = (m_{\pi^{\rm ave}}, m_{K^{\rm ave}}, m_{\Xi^{\rm ave}})$, 
the lattice cutoff, $a^{-1}$, increases by $\sim 1.1\%$
and
$m_{ud} (m_{s})$ decreases by $\sim 0.6 (4.5)\%$, 
compared to our main results.
In the case of
the definition with 
$(m_{\pi}, m_K, m_{\Xi}) = (m_{\pi^0}, m_{K^0}, m_{\Xi^0})$,
the lattice cutoff %, $a^{-1}$, 
increases by $\sim 0.03\%$
and
$m_{ud} (m_{s})$ decreases by $\sim 3.1 (1.6)\%$, 
compared to our main results.
Note again that these are associated with 
inevitable ambiguities of the physical point
in $(2+1)$-flavor lattice QCD~\cite{FlavourLatticeAveragingGroupFLAG:2021npn},
and 
it is desirable to perform 
lattice QCD + QED simulations 
with the isospin breaking effects %taken into account
in future.

%%%%%%%%%%%%%%%%%%%%%%%%%%%%%%%%%%%%%%%%%
%%summary table for hadron masses in physical unit %
\begin{table}[hbtp]
  \caption{
  Summary table for the masses of stable hadrons in physical units.
  The first and second parentheses denote the statistical 
  and systematic errors, respectively.
  The experimental values correspond to isospin-averaged ones
  taken from Particle Data Group (PDG)~\cite{Workman:2022ynf}.
  } 
  \label{table:HadronMassPhysical}
  \centering
  \begin{tabular}{ccc}
    \hline
    hadron & experiment [MeV] & lattice [MeV] \\
    \hline %\hline
    $\pi$  & 138.04 & $137.1(0.3)(^{+0.0}_{-0.2})$ \\
    $K$  & 495.64 & $501.8(0.3)(^{+0.0}_{-0.7})$ \\
    \hline 
    $N$  & 938.92 & $939.7(1.8)(^{+0.2}_{-1.7})$ \\
    $\Lambda$  & 1115.68 & $1121.4(3.6)(^{+0.5}_{-3.7})$ \\
    $\Sigma$  & 1193.15 & $1202.5(5.6)(^{+0.3}_{-4.0})$ \\
    $\Xi$  & 1318.29 & $1320.7(2.1)(^{+1.5}_{-1.7})$ \\
    \hline
    $\Omega$  & 1672.45 & input \\
    \hline
  \end{tabular}
\end{table}
%%%%%%%%%%%%%%%%%%%%%%%%%%%%%%%%%%%%%%%%%

%%%%%%%%%%%%%%%%%%%%%%%%%%%%%%%%%%%%%%
%%%%%%%%%%%%%%%%%%%%%%%%%%%%%%%%%%%%%%
\section{Summary and concluding remarks}\label{sec:summary}
%%%%%%%%%%%%%%%%%%%%%%%%%%%%%%%%%%%%%%
%%%%%%%%%%%%%%%%%%%%%%%%%%%%%%%%%%%%%%

In this paper,
we have reported the scale setting and hadronic properties 
for a new set of lattice QCD gauge configurations generated by the HAL QCD Collaboration
which is named as ``HAL-conf-2023".
The purpose of this new generation is to enable lattice simulations
%(i) 
at the physical point, 
%(ii) 
on a large lattice volume and 
%(iii) 
with a large number of ensembles.
We employed $(2+1)$-flavor nonperturbatively ${\cal O}(a)$ improved Wilson fermions 
with stout smearing and the Iwasaki gauge action at $\beta = 1.82$.
The size of the lattice is $96^4$, corresponding to $(8.1 {\rm fm})^4$ in  physical units.
%Utilizing the supercomputer Fugaku,
We generated as many as 8,000 trajectories at the physical point
using the simulation parameters employed
by the PACS Collaboration~\cite{Ishikawa:2018jee, PACS:2019ofv}.

We showed the MD history of the generation
and basic properties of the configurations including the plaquette value, the topological charge distribution and their auto-correlation times.
We observed that the MD evolution was stable, and the auto-correlation times 
%of the configurations 
are short even for the topological charge, 
indicating that this configuration set is well spread in the configuration space.

We performed the scale setting using the $\Omega$ baron mass
as a reference scale. The systematic error has been carefully examined by 
the operator dependence of the correlation function of $\Omega$,
and the value of the lattice cutoff is determined as
$a^{-1} =2338.8(1.5)(^{+0.2}_{-3.0})$ MeV.

We calculated the correlation functions of hadrons
and obtained the physical results of single hadron spectra,
quark masses and decay constants of pseudoscalar mesons
in the light quark sector.
The masses of the pseudoscalar mesons are found to be
$(m_\pi, m_K) = (
137.1(0.3)(^{+0.0}_{-0.2}), 
501.8(0.3)(^{+0.0}_{-0.7}))$ MeV, 
which are sufficiently close to the experimental values 
considering the uncertainties of the isospin breaking.
The masses of other stable hadrons also agree 
with the experimental values within a sub-percent level.
The results of the quark masses and decay constants are compared to
those from PACS10 configuration with the same simulation parameter~\cite{Ishikawa:2018jee, PACS:2019ofv}
and those of FLAG average~\cite{FlavourLatticeAveragingGroupFLAG:2021npn},
and are found to be consistent within uncertainties.
These results show that 
this configuration set, HAL-conf-2023,
serves as good physical point ensembles
to be used for future lattice QCD calculations for various physical quantities.
One of the notable utilities of HAL-conf-2023
is the physical point simulations of hadron interactions,
which are currently in progress and will be reported elsewhere.

Finally, 
we remark several uncertainties to be investigated in the future.
The present calculation is performed at only one lattice spacing 
and the continuum limit is yet to be taken.
While the agreements of our results with experimental values
as well as with the FLAG average values (in the continuum limit)
indicate that the scaling violation is suppressed for this configuration set,
explicit calculations at finer lattice spacings would be desirable. 
In addition, the ``physical point" in our $(2+1)$-flavor lattice QCD simulation is defined
only up to the uncertainties of the isospin breaking effects.
The vacuum polarization effect of the charm quark has not been taken into account, either.
In the future, it is desirable to perform 
$(2+1+1)$, $(1+1+1)$ and $(1+1+1+1)$-flavor QCD (+ QED) simulations 
to fully control the systematics in lattice calculations.

%%%%%%%%%%%%%%%%%%%%%%%
%%%%%%%%%%%%%%%%%%%%%%%
\begin{acknowledgments}
%%%%%%%%%%%%%%%%%%%%%%%
%%%%%%%%%%%%%%%%%%%%%%%

We would like to thank 
Drs.
K.-I.~Ishikawa, 
I.~Kanamori and
Y.~Nakamura 
for the simulation code and 
helpful discussions for configuration generation,
and 
Dr. N.~Ukita for useful information
on PACS configurations~\cite{Ishikawa:2018jee, PACS:2019ofv}.
We thank the authors of 
Bridge++~\cite{Ueda:2014rya} and
cuLGT code~\cite{Schrock:2012fj} used for this study.
%used for 
%the the gauge fixing.
We thank other members of the HAL QCD Collaboration for fruitful discussions.
The numerical lattice QCD simulations have been performed on 
the supercomputer Fugaku at RIKEN
and Cygnus supercomputer at University of Tsukuba.
We thank ILDG/JLDG \cite{ldg, Amagasa:2015zwb}, which serves as essential infrastructure in this study.
This work was partially supported by
HPCI System Research Project
(hp200130, hp210165, hp220174, hp230207, hp240213, hp220066, hp230075, hp240157, hp210212, hp220240),
the JSPS
(Grants No. 
JP18H05407, % cluster shingakujutu 
JP18H05236, % hatsuda kiban-s
JP19K03879, % doi kiban-c
JP21H05190, % itou  Transformative Research Areas (A) 
JP23H05439), % doi kiban-s
JST PRESTO Grant Number JPMJPR2113, % itou
``Priority Issue on Post-K computer'' (Elucidation of the Fundamental Laws and Evolution of the Universe), 
``Program for Promoting Researches on the Supercomputer Fugaku'' (Simulation for basic science: from fundamental laws of particles to creation of nuclei) 
and (Simulation for basic science: approaching the new quantum era)
(Grants No. JPMXP1020200105, JPMXP1020230411), 
and Joint Institute for Computational Fundamental Science (JICFuS).

\end{acknowledgments}
%%%%%%%%%%%%%%%%%%%%%%%
%%%%%%%%%%%%%%%%%%%%%%%
%%%%%%%%%%%%%%%%%%%%%%%
\appendix

%%%%%%%%%%%%%%%%%%%%%%%%
\section{The ratio-difference method for quark masses}
\label{sec:ratio-diff}

In order to obtain physical quark masses,
the so-called ratio-difference method was developed by the BMW Collaboration~\cite{BMW:2010skj}.
In this method, AWI and VWI bare quark masses are combined
in a different way than the one used in the main text 
(Sec.~\ref{subsec:meson-strategy}),
considering that physical quark masses are obtained by
additive and multiplicative (only multiplicative) renormalization 
for VWI (AWI) masses for Wilson-type fermions.

While their concept and the derivation were well-described in~\cite{BMW:2010skj},
we find that the obtained formula in Ref.~\cite{BMW:2010skj} has to be corrected.
This is because they performed the ${\cal O}(ma)$ improvement for AWI masses in each flavor
(Eq.~(11.6) in Ref.~\cite{BMW:2010skj}), while such improvement is valid 
only for flavor non-singlet sectors~\cite{Bhattacharya:2005rb}.
In this appendix, we give a formula derived with the correct ${\cal O}(ma)$ improvement.
In addition, we take into account the ${\cal O}(g_0^4)$ terms which are neglected in Ref.~\cite{BMW:2010skj}.

We define the AWI bare quark masses as
\beq
m_{f}^{\rm AWI}+m_g^{\rm AWI} = \frac{\langle 0| \partial_4 A_4 | PS \rangle  }{\langle 0 | P | PS \rangle }.
\eeq
and 
the VWI bare quark masses
\footnote{
In the literature, it is usually called just as {\it "bare quark masses"}.
}
\beq
m_f^{\rm VWI} = \frac{1}{a}\left(\frac{1}{2\kappa_f} - \frac{1}{2\kappa_c}\right) .
\eeq
Here, ${\cal O}(a)$ improvement of the axial-current $A_4 \rightarrow A_4 + a c_A\partial_4 P$ is implicitly understood,
and the flavor indices $f, g$ should be taken such that the axial-current is flavor non-singlet.

The physical quark masses $\hat{m}_f$ obtained by the ${\cal O}(ma)$ improvement and renormalization for 
the AWI or VWI bare quark masses are given by~\cite{Bhattacharya:2005rb}

%%%%%%%%%%%%%%%%%%%%%
\begin{eqnarray}
  \hat{m}_{fg} &=& \frac{Z_A}{Z_P} \left[
    \frac{1 + a b_A m_{fg} + a \bar{b}_A {\rm Tr} M}
    {1 + a b_P m_{fg} + a \bar{b}_P {\rm Tr} M}
    \right]
  m_{fg}^{{\rm AWI}} 
  + {\cal O}(a^2), 
  \label{eq:AWI-mass-phys}
\end{eqnarray}
where
$m'_{fg} = (m'_f + m'_g)/2$ with $m'$ denoting $\hat{m}, m$ or $m^{\rm AWI}$,
%%%%%%%%%%%%%%%%%%%%%
%$\hat{m}_{fg} = (\hat{m}_f + \hat{m}_g)/2$,
%$m_{fg}^{\rm AWI} = (m_f^{\rm AWI} + m_g^{\rm AWI})/2$,
%$m_{fg}^{\rm VWI} = (m_f^{\rm VWI} + m_g^{\rm VWI})/2$,
%
and
%
%%%%%%%%%%%%%%%%%%%%%
\begin{eqnarray}
  \hat{m}_{f} 
  &=& Z_m \left\{
  \left[ m_f + (r_m - 1)\frac{{\rm Tr}M}{N_f}\right] 
  \right. \nn
  && \qquad \left.
  + a \left[ b_m m_f^2 + \bar{b}_m m_f {\rm Tr}M 
    + (r_m d_m - b_m)\frac{{\rm Tr}(M^2)}{N_f}  
    + (r_m \bar{d}_m - \bar{b}_m)\frac{({\rm Tr}M)^2}{N_f}
    \right]
  \right\} 
  + {\cal O}(a^2) . \nn
\label{eq:VWI-mass-phys-each}
\end{eqnarray}
%%%%%%%%%%%%%%%%%%%%%

In the following, we employ $m^{\rm VWI}$ for the bare mass $m$ 
appearing in the ${\cal O}(ma)$ improvement terms
as is usually the case in the literature~\cite{Bhattacharya:2005rb},
and the improvement coefficients, $b_X, \bar{b}_X, d_X, \bar{d}_X (X=A, P, m)$, 
are defined accordingly.

We now derive the formula for the ratio-difference method.
For simplicity, we consider $N_f = 2+1$ QCD hereafter,
with flavor indices $(u,d,s)$ labeled by $(1,2,3)$.
The formula for other flavor cases can be obtained similarly.

We begin with the following definitions and trivial equalities,
\begin{eqnarray}
  d^{\rm imp} &\equiv& \frac{1}{Z_m} (a \hat{m}_{s} - a \hat{m}_{ud}) \\
  d &\equiv& a m_{s}^{{\rm VWI}} - a m_{ud}^{{\rm VWI}} \\
  r^{\rm imp} &\equiv& \frac{\hat{m}_{s}}{\hat{m}_{ud}} \\
  r &\equiv& \frac{m_{s}^{{\rm AWI}}}{m_{ud}^{{\rm AWI}}}
  = 2\frac{m_{13}^{{\rm AWI}}}{m_{12}^{{\rm AWI}}} -1 \\
  a \hat{m}_{ud} &=& Z_m \frac{d^{\rm imp}}{r^{\rm imp}-1} \\
  a \hat{m}_{s}  &=& Z_m \frac{r^{\rm imp}d^{\rm imp}}{r^{\rm imp}-1} 
\end{eqnarray}
From Eqs.~(\ref{eq:AWI-mass-phys}) and (\ref{eq:VWI-mass-phys-each}),
\begin{eqnarray}
  d^{\rm imp} &=&
  (a m_s^{{\rm VWI}} - a m_{ud}^{{\rm VWI}}) \times 
  \left[
    1 +
    a b_m (m_s^{{\rm VWI}} + m_{ud}^{{\rm VWI}})
    + a \bar{b}_m {\rm Tr}M^{\rm VWI}
   + {\cal O}(a^2) 
    \right] 
   \label{eq:diff1}
\end{eqnarray}
\begin{eqnarray}
  \frac{\hat{m}_{13}}{\hat{m}_{12}} &=& \frac{1}{2} r^{\rm imp} + \frac{1}{2} \nn
  &=&
  \frac{m_{13}^{{\rm AWI}}}{m_{12}^{{\rm AWI}}}
  \left[
    1 + a (b_A - b_P) (m_{13}^{{\rm VWI}} -  m_{12}^{{\rm VWI}})
    + {\cal O}(a^2)
    \right]
   \nn
  &=&
  \frac{r+1}{2}
  \left[
    1 + (b_A - b_P) \frac{1}{2} d
    + {\cal O}(a^2) 
    \right] 
  \label{eq:ratio1}
\end{eqnarray}

The next step is to express 
$(m_s^{{\rm VWI}} + m_{ud}^{{\rm VWI}})$
and
${\rm Tr}M^{\rm VWI} = (m_s^{{\rm VWI}} + 2m_{ud}^{{\rm VWI}})$
which appear in $d^{\rm imp}$ by some functions of $r, d$.
For this purpose, we use Eqs.~(\ref{eq:ratio1}) and (\ref{eq:VWI-mass-phys-each}) to show that
\begin{eqnarray}
  \frac{\hat{m}_s}{\hat{m}_{ud}} 
%  &=& \frac{m_{s}^{{\rm AWI},I}}{m_{ud}^{{\rm AWI},I}} + {\cal O}(a) \nn
  &=& r + {\cal O}(a) 
  \label{eq:ratio2}
\end{eqnarray}
and
\begin{eqnarray}
  \frac{\hat{m}_s}{\hat{m}_{ud}} &=&
  \frac{m_{s}^{{\rm VWI}}  + (r_m-1)\frac{1}{3}(m_{s}^{{\rm VWI}} + 2 m_{ud}^{{\rm VWI}})}
       {m_{ud}^{{\rm VWI}} + (r_m-1)\frac{1}{3}(m_{s}^{{\rm VWI}} + 2 m_{ud}^{{\rm VWI}})} 
  + {\cal O}(a) .
  \label{eq:ratio3}
\end{eqnarray}

Equating Eqs.~(\ref{eq:ratio2}) and (\ref{eq:ratio3}),
we obtain
\begin{eqnarray}
  \frac{m_{s}^{{\rm VWI}}}{m_{ud}^{{\rm VWI}}}
  &=&
  \frac{(2r_m+1) r - 2(r_m-1)}{(r_m+2) - (r_m-1)r} + {\cal O}(a) 
\end{eqnarray}
and thus
\begin{eqnarray}
  a m_s^{{\rm VWI}} + a m_{ud}^{{\rm VWI}}
  &=&
  (a m_s^{{\rm VWI}} - a m_{ud}^{{\rm VWI}}) \cdot
  \frac{m_s^{{\rm VWI}}/m_{ud}^{{\rm VWI}} + 1}{m_s^{{\rm VWI}}/m_{ud}^{{\rm VWI}} - 1} \nn
  &=&
  d 
  \left[
  \frac{3(r+1) + (r_m-1)(r-1)}{3r_m (r-1)} + {\cal O}(a)
  \right]
\end{eqnarray}
\begin{eqnarray}
  a m_s^{{\rm VWI}} + 2 a m_{ud}^{{\rm VWI}}
  &=&
  (a m_s^{{\rm VWI}} - a m_{ud}^{{\rm VWI}}) \cdot
  \frac{m_s^{{\rm VWI}}/m_{ud}^{{\rm VWI}} + 2}{m_s^{{\rm VWI}}/m_{ud}^{{\rm VWI}} - 1} \nn
  &=&
  d
  \left[
  \frac{r+2}{r_m (r-1)} + {\cal O}(a) 
  \right]
\end{eqnarray}

To summarize, the formula for the ratio-difference method is
\begin{eqnarray}
%\left\{
%\begin{gathered}
  a \hat{m}_{ud} &=& Z_m \frac{d^{\rm imp}}{r^{\rm imp}-1} \\
  a \hat{m}_{s}  &=& Z_m \frac{r^{\rm imp}d^{\rm imp}}{r^{\rm imp}-1} \\
  d^{\rm imp} 
%  &\equiv& \frac{1}{Z_m} (a \hat{m}_{s} - a \hat{m}_{ud}) \nn
  &=&
  d \cdot
  \left[
    1 +
    b_m
    d \cdot
    \frac{3(r+1) + (r_m-1)(r-1)}{3r_m (r-1)} 
    + \bar{b}_m
    d \cdot \frac{r+2}{r_m (r-1)} 
    + {\cal O}(a^2)
    \right]  \\
  r^{\rm imp} 
%  &\equiv& \frac{\hat{m}_{s}}{\hat{m}_{ud}} \nn
  &=& 
  r 
  \left[
    1 + (b_A - b_P) d\cdot \frac{r+1}{2r} 
    + {\cal O}(a^2)
    \right]   \\
  d 
%  &\equiv& a m_{s}^{{\rm VWI}} - a m_{ud}^{{\rm VWI}} 
  &=& \frac{1}{2\kappa_s} - \frac{1}{2\kappa_{ud}} \\
  r 
%  &\equiv& \frac{m_{s}^{{\rm AWI}}}{m_{ud}^{{\rm AWI}}}
  &=& 2\frac{m_{13}^{{\rm AWI}}}{m_{12}^{{\rm AWI}}} -1 ,
%\end{gathered}
%\right.
\end{eqnarray}
where $d^{\rm imp}$ and $r^{\rm imp}$ are different 
from the original formula given by the BMW Collaboration 
(Eqs. (11.14) and (11.15) in Ref.~\cite{BMW:2010skj}).

We note that, in numerical calculations for quark masses
by the BMW Collaboration~\cite{BMW:2010skj, BMW:2013fzj},
the tree-level improvement coefficients are used in practice, where 
$(b_A - b_P) = \bar{b}_A = \bar{b}_P = \bar{b}_m = 0, r_m = 1$.
In such a case, the differences in the formula become irrelevant
and their numerical results remain intact.

%%%%%%%%%%%%%%%%%%%%%%%%%%%5

\section{Comparison with PACS10 study}\label{sec:PACS10}

We present
a comparison of physical point simulations
with our  HAL-Conf-2023 configurations
and PACS10 configurations
by the PACS collaboration~\cite{Ishikawa:2018jee, PACS:2019ofv},
where
the same action with the same parameters
such as hopping parameters $\kappa$
%$(\kappa_{\rm ud},\kappa_{\rm s})=(0.126117,0.124902)$ 
(except for the lattice volume)
are employed in the generation.
More specifically,
we find that
there is a deviation of $\simeq 1\%$ 
(corresponding to 4.1$\sigma$)
for the obtained values of the lattice cutoff,
$a^{-1}_{\rm HAL}=2338.8(1.5)\left(^{+0.2}_{-3.0}\right)$ MeV in our study
and
$a^{-1}_{\rm PACS10}=2316.2(4.4)$ MeV in the PACS10 study,
and we examine the origin of the difference.

First, we consider the effect of the difference of 
the lattice volume,
where $N_s^4=96^4$ ($L = N_s a \simeq 8.1$ fm)
in  HAL-Conf-2023
and $N_s^4=128^4$ ($L \simeq 10.8$ fm)
in PACS10 configurations.
In the study of the PACS collaboration%
~\cite{Ishikawa:2018jee, PACS:2019ofv},
configurations with
$N_s^4=64^4$ ($L \simeq 5.4$ fm)
with the same hopping parameters
are also generated,
and the PS meson masses and the PCAC quark masses are found to be different 
between $N_s=64$ and $N_s=128$
due to the finite volume effect on the critical kappa $\kappa_c$.
On this point, 
we find that 
our PS meson masses and PCAC masses in lattice units
are consistent with those of PACS10 as shown in Sec.~\ref{subsec:PS-PCAC-masses}. 
%so that we conclude that
This indicates that
$N_s=96$ is large enough to be compared
directly with $N_s=128$.
In addition, 
since the PCAC masses (and baryon masses used for the scale setting) 
in lattice units become larger 
with a smaller volume in the PACS study,
our setup with $N_s=96$ might underestimate
the cutoff value $a^{-1}$ compared to $N_s=128$ 
due to the finite volume effect on $\kappa_c$, 
if any.
This is the opposite effect of the deviation
between $a^{-1}_{\rm HAL}$ and $a^{-1}_{\rm PACS10}$.
As further consideration,
even when finite volume effect on $\kappa_c$
is negligible, there could also exist finite volume effect 
on baryon masses through, e.g., meson cloud contributions. 
However, such an effect is considered to be negligible 
for our large spatial volume, 
as is estimated by ChPT in Sec.~\ref{subsec:baryon-results}
and
is numerically demonstrated
in the PACS study for the comparison of baryon spectra
between $N_s=128$ and $N_s=64$ (with reweighted $\kappa$ for the latter
to compensate for the finite volume effect on $\kappa_c$).
With these considerations, 
we conclude that the difference of the lattice volume
is not the origin of that between
$a^{-1}_{\rm HAL}$ and $a^{-1}_{\rm PACS10}$.

Second, we consider the 
difference for the definition of the ``physical point".
In the PACS10 study,
quark masses are slightly reweighted from the simulation point
using the experimental values of 
$(m_{\pi^0}, m_{K^0})$
with the lattice scale determined by $m_{\Xi^0}$.
On the other hand, we do not use the reweighting since 
the spectrum on the simulation point is 
sufficiently close to our physical point target defined by 
the experimental spectra 
with the isospin average, $(m_{\pi^{\rm ave}}, m_{K^{\rm ave}})$,
with the lattice scale determined by $m_{\Omega}$.
Among several differences given above, the effect of reweighting on $m_{\Xi^0}$
in PACS10 study is much smaller 
than the difference between $a^{-1}_{\rm HAL}$ and $a^{-1}_{\rm PACS10}$~\cite{PACS:2019ofv}
and can be neglected.
Also, the choice of baryon species for the scale setting is not an issue
within our study, since the lattice cutoff determined from $m_\Omega$ is consistent with
that determined from $m_{\Xi^{\rm ave}}$.
However, the choice of experimental value for $m_{\Xi}$ does matter.
In fact, the experimental value of $m_{\Xi^0}$
is about $0.3\%$ smaller than $m_{\Xi^{\rm ave}}$
and thus can explain 
a part of the $1\%$ deviation
between $a^{-1}_{\rm HAL}$ and $a^{-1}_{\rm PACS10}$.
Note that this difference is unavoidable uncertainty
for the $N_f = 2+1$ lattice QCD (without QED) simulations, 
and there is no rigorous theoretical argument
which one to choose~\cite{FlavourLatticeAveragingGroupFLAG:2021npn}.

Finally, we examine the origin of 
the remaining $0.7\%$ deviation for the lattice cutoff.
It turns out that 
the mass of $\Xi$ in lattice units, 
obtained on the lattice at the simulation point, 
has 0.5\% deviation ($2.6 \sigma$)
as
$(a m_\Xi)_{\rm HAL} = 0.56469(74)\left(^{+58}_{-0}\right)$
and 
$(a m_\Xi)_{\rm PACS10} = 0.56742(50)$.
In these calculations,
the correlator with 
the wall-source and point-sink operator is employed in our study,
while 
the smeared-source and point-sink operator is employed in the PACS10 study.
In order to further study the dependence on the source operator,
we additionally calculate with smeared-source and point-sink operator
with two smearing parameter sets, $l=1$ (narrow) and $l=2$ (broad) given in Sec.~\ref{subsec:baryon-strategy}.

\begin{figure}[htbp]
 \begin{center}
        \includegraphics[keepaspectratio, scale=0.55]{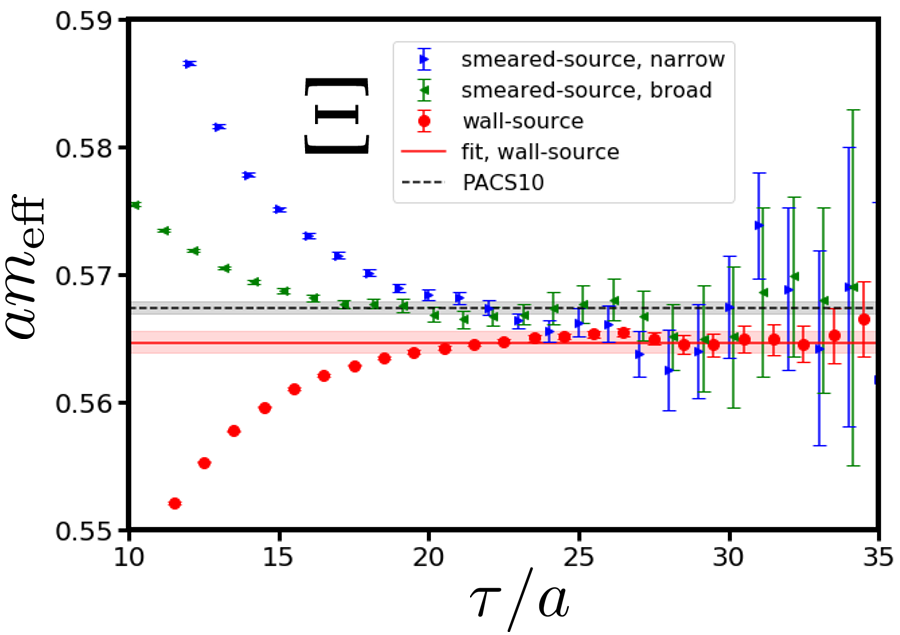}
    \caption{
    Effective mass plot for the $\Xi$ baryon in lattice units with two methods: the point-sink and wall-source operators (red circles) and the point-sink and smeared-source operators 
    (blue (green) triangles for narrow (broad) smearing function).
    The red horizontal line denotes the fit results with the wall-source and point-sink operator, and the red band denotes the statistical and systematic errors combined in quadrature. The black dashed line with a band denotes the fit result
    of the PACS10 study, obtained from the 
    smeared-source and point-sink operator at the simulation point~\cite{PACS:2019ofv}.
}\label{fig:XiPACS10}
\end{center}
  \end{figure}

The effective masses in our calculations are shown in Fig.\ref{fig:XiPACS10}.
The red circles denote the results
with the wall-source, % and point-sink operator
and the 
blue (green) triangles denote 
those with the narrow (broad) smeared-source. 
The red horizontal line with a band denotes our fit result
obtained from the wall-source
with statistical and systematic errors added in quadrature.
As Fig.\ref{fig:XiPACS10} shows, 
all results are consistent 
with the red band
in the large $\tau$ region.
In particular,
while the wall-source data tend to converge from the lower side,
our smeared-source data tend to converge from the upper side,
indicating that our identification of the ground state plateau is reliable.

In Fig.\ref{fig:XiPACS10},
we also show the result of the PACS10 study
for comparison,
where 
the black dashed line with a band  
is their fit result with their error
obtained from the smeared-source at the simulation point.
There is a clear deviation from our results as mentioned before.
We note that the smearing functions are slightly different
between our study and the PACS10 study,
and the black dashed line does not correspond to the
fit of our smeared-source data (blue and green triangles).
Having said that, a comparison implies that,
if one analyzes a correlator with only one specific choice of the operator,
one may easily misidentify a pseudo-plateau structure at early $\tau$
as a signal of the ground state saturation, 
which leads to an incorrect result for the spectrum.

As discussed above, our analysis for $m_\Xi$ passed the consistency check
between different choices of the source operator.
In addition, our final result for the lattice cutoff is determined
from $m_\Omega$, where a more stringent and higher-precision 
cross check is performed 
between the result from the wall-source and point-sink operator
and that from the variational method with the $2\times 2$ smeared operators
(See Sec.~\ref{subsec:baryon-results}).
Therefore, we conclude that the reliability of our determination of the lattice cutoff
is well established.

%%%%%%%%%%%%%%%%%%%%%%%%%%%%%%%
 %%%%%%%%%%%%%%%%%%%%%%%%%%%%%%%%%%%%%

\section{The effective masses of unstable hadrons}
\label{sec:unstable-app}
\begin{figure}[htbp]
 \begin{center}
        \includegraphics[keepaspectratio, scale=0.4]{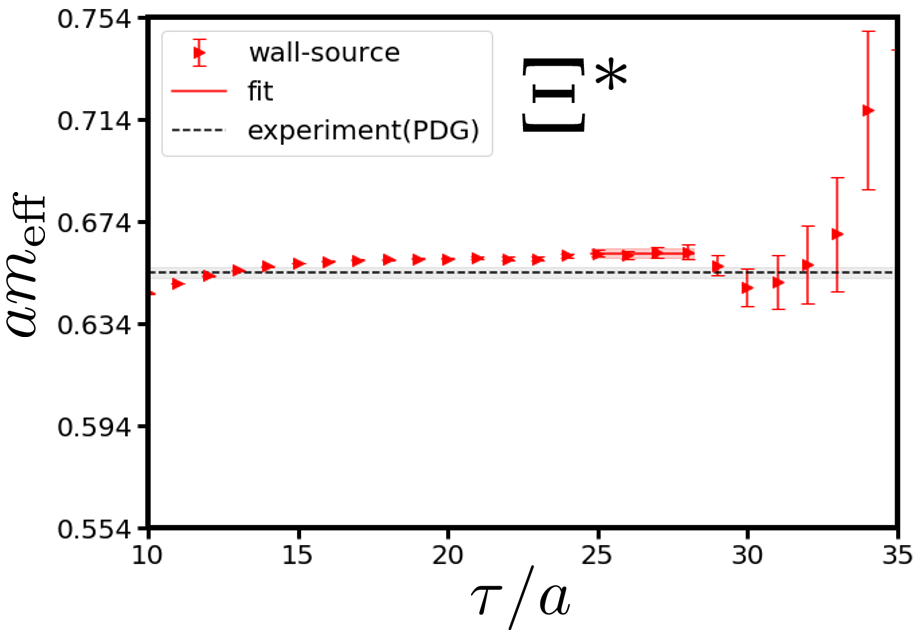}
        \includegraphics[keepaspectratio, scale=0.4]{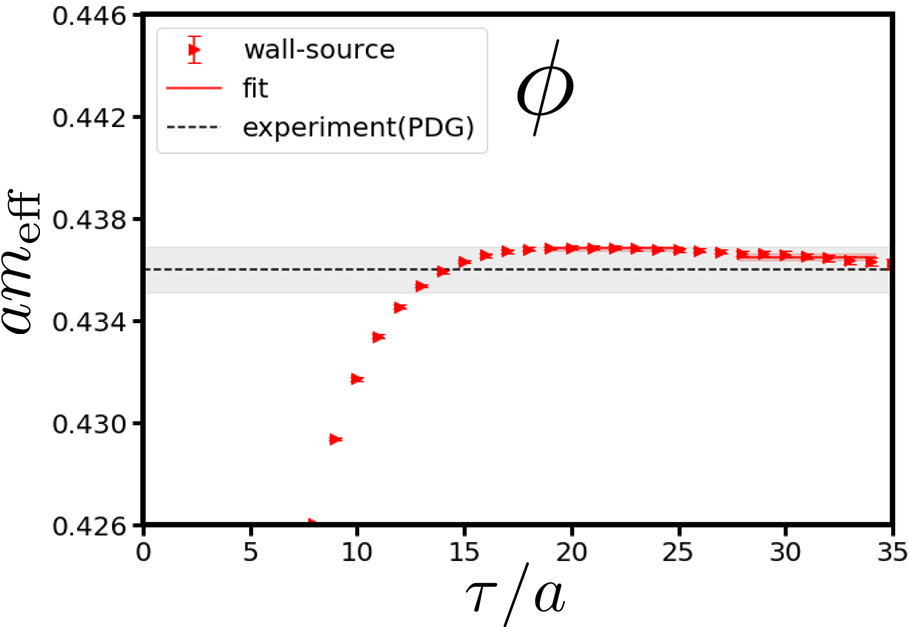} \\
        \includegraphics[keepaspectratio, scale=0.4]{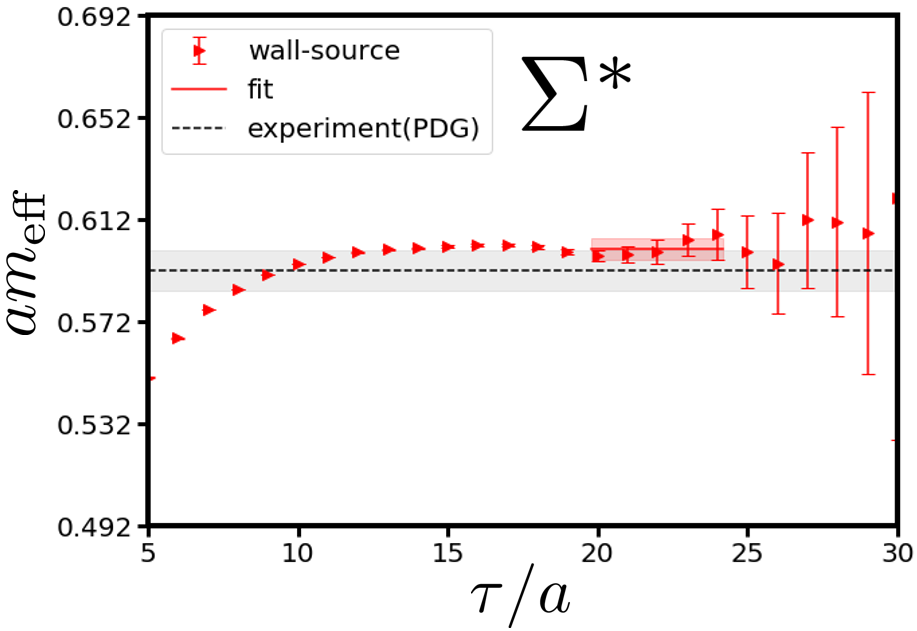}
        \includegraphics[keepaspectratio, scale=0.4]{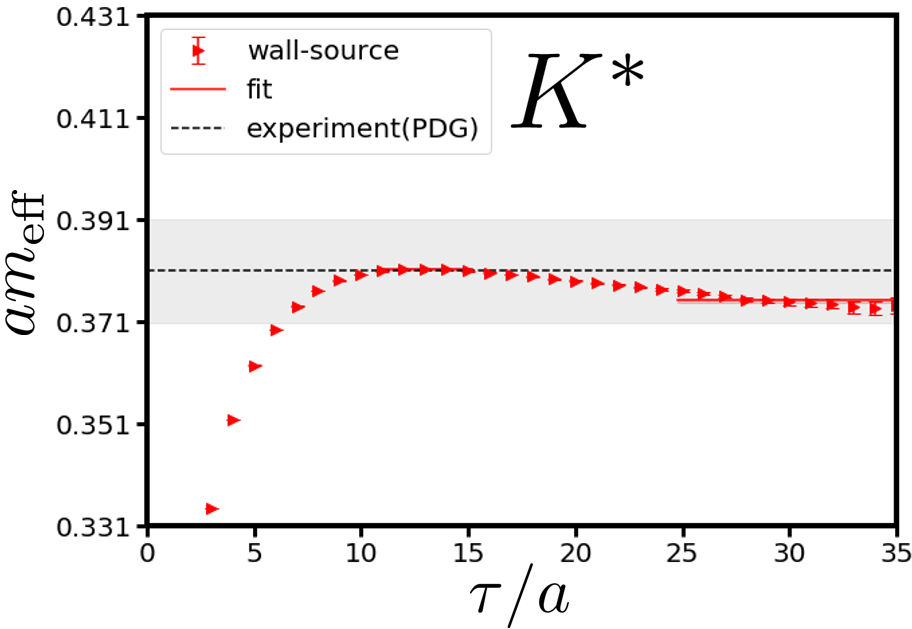}\\
        \includegraphics[keepaspectratio, scale=0.4]{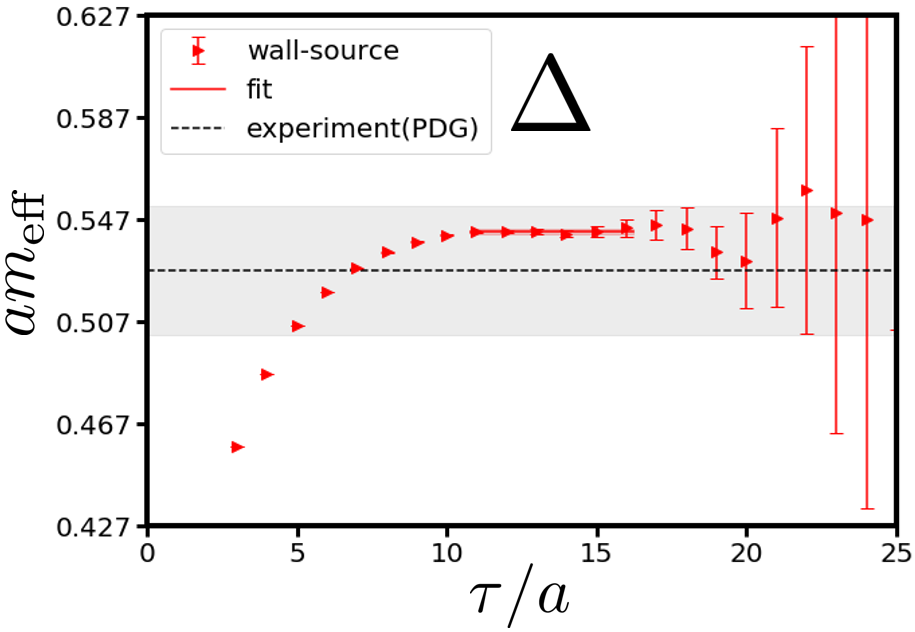}
        \includegraphics[keepaspectratio, scale=0.4]{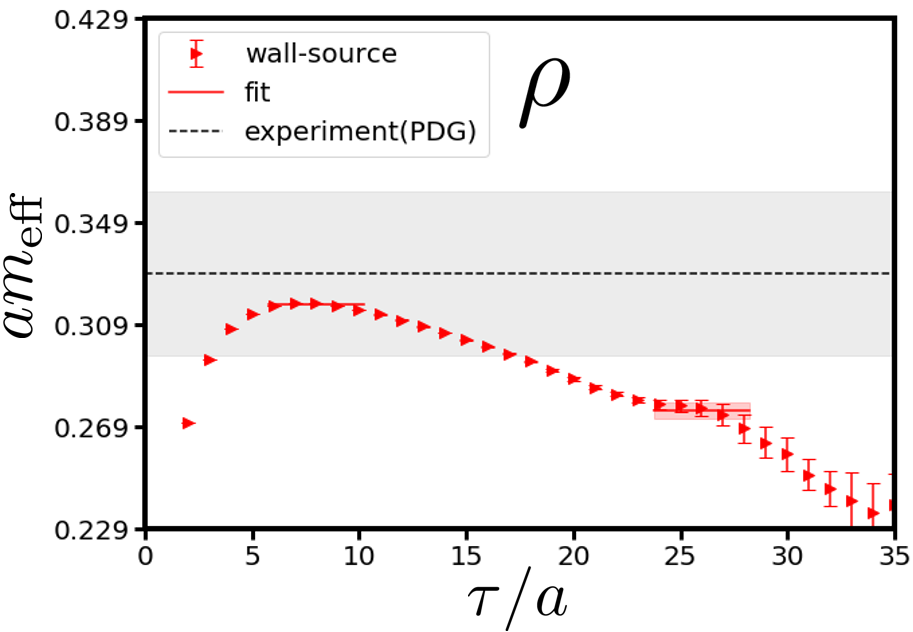} 
    \caption{
    Effective masses for low-lying resonant
    decuplet baryons and vector mesons 
    with point-sink and wall-source correlators in lattice units.
    The red lines and red bands denote
    the fit results and the statistical errors, respectively. 
    The black dotted lines and the gray bands 
    denote the experimental values
    of masses and widths, respectively.
}\label{fig:resonance}
\end{center}
  \end{figure}

We show the effective mass plots for the unstable hadrons (resonances)
discussed in Sec.~\ref{sec:unstable}.
In Fig.~\ref{fig:resonance}, 
the results are shown by red points.
Shown together by red lines with red bands
are the results of the single-state fit of the correlators
with statistical errors.
Systematic errors are estimated by the fit range dependence.
In some hadrons, we do not see clear single plateaux,
and we adopt two separate fit ranges for possible plateaux
as explicitly shown in Fig.\ref{fig:resonance}.

%--- 5 each baryon effective mass plots --- 図だけappendixに動かす%
%%%%%%%%%%%%%%%%%%%%%%%%%%%%%%%
 %%%%%%%%%%%%%%%%%%%%%%%%%%%%%%%%%%%%%

%%%%%%%%%%%%%%%%%%%%%%%
%%%%%%%%%%%%%%%%%%%%%%%
%%%%%%%%%%%%%%%%%%%%%%%
\bibliographystyle{utphys}
\bibliography{latticeQCD}
%%%%%%%%%%%%%%%%%%%%%%%
%%%%%%%%%%%%%%%%%%%%%%%
%%%%%%%%%%%%%%%%%%%%%%%

\end{document}